\documentclass[twocolumn]{aastex61}

\usepackage{natbib,aas_macros} 
\usepackage{hyperref}
\usepackage{graphicx}	
\usepackage{amsmath,amssymb} 
\usepackage{xspace}
\usepackage{enumerate}
\usepackage{fp} 
\usepackage{lineno} 

 \newcommand{\Iextinction}{0.031}
 \newcommand{\Vextinction}{0.051}
 \newcommand{\trgblum}{-3.95}
 \newcommand{\trgblumstaterr}{0.03} 
 \newcommand{\trgblumsyserr}{0.05}

 \newcommand{\trgbobsval}{27.371}
 \newcommand{\trgbobsvalstaterr}{0.03} 
 \newcommand{\trgbobsvalsyserr}{0.01}

 \newcommand{\wavgdistance}{$\langle\mu_0\rangle=31.30\pm0.03$~mag\xspace}

 \FPeval\trgbobsvalrounded{round(\trgbobsval,2)}
 \FPeval\trgbredcorrval{round(\trgbobsval-\Iextinction,2)}
 \FPeval\truetrgbdmod{round(\trgbobsval-\Iextinction-\trgblum,2)}
 \FPeval\truetrgbdmodMpc{round(10^(\truetrgbdmod/5)/100000,1)}

 \FPeval\dmodcombinedstaterr{ round( (\trgbobsvalstaterr^2+\trgblumstaterr^2)^0.5,2) }
 \FPeval\dmodcombinedsyserr{ round( (\trgbobsvalsyserr^2+\trgblumsyserr^2)^0.5,2) }

 \FPeval\truetrgbdmodMpcupperrdiststat{ 10^( (\truetrgbdmod+\dmodcombinedstaterr) /5)/100000 }
 \FPeval\truetrgbdmodMpclowerdiststat{ 10^( (\truetrgbdmod-\dmodcombinedstaterr) /5)/100000 }
 \FPeval\truetrgbdmodMpcstaterr{ round( 0.5*(\truetrgbdmodMpcupperrdiststat - \truetrgbdmodMpclowerdiststat) ,1) }
 \FPeval\truetrgbdmodMpcupperrdistsys{ 10^( (\truetrgbdmod+\dmodcombinedsyserr) /5)/100000 }
 \FPeval\truetrgbdmodMpclowerdistsys{ 10^( (\truetrgbdmod-\dmodcombinedsyserr) /5)/100000 }
 \FPeval\truetrgbdmodMpcsyserr{ round( 0.5*(\truetrgbdmodMpcupperrdistsys - \truetrgbdmodMpclowerdistsys) ,1) }
 
 \newcommand{\trgbobsvalwerr}{$\mathrm{F814W}=\trgbobsvalrounded\pm \trgbobsvalstaterr_{stat}\pm \trgbobsvalsyserr_{sys}~\mathrm{mag}$\xspace}
 \newcommand{\trgbredcorrvalwerr}{$\mathrm{F814W}=\trgbredcorrval\pm \trgbobsvalstaterr_{stat}\pm \trgbobsvalsyserr_{sys}~\mathrm{mag}$\xspace}
 \newcommand{\truetrgbdmodwerr}{$\mu_0 = \truetrgbdmod \pm\dmodcombinedstaterr_{stat} \pm\dmodcombinedsyserr_{sys}~\mathrm{mag}$\xspace}
 \newcommand{\truetrgbdmodMpcwerr}{$D = \truetrgbdmodMpc\pm\truetrgbdmodMpcstaterr_{stat}\pm\truetrgbdmodMpcsyserr_{sys}$ Mpc\xspace}
 \newcommand{\trgblumwerr}{$M_{I}^\mathrm{TRGB}=\trgblum\pm\trgblumstaterr_{stat}\pm\trgblumsyserr_{sys}$\xspace}

 \newcommand{\ngc}{NGC\,1365\xspace}
 \newcommand{\sne}{SNe~Ia\xspace}
 \newcommand{\cchp}{\citetalias{cchp2proposal}\xspace} 
 \newcommand{\ho}{$H_0$\xspace}
 \newcommand{\hst}{\emph{HST}\xspace}
 
 \newcommand{\sigmasmooth}{$\sigma_s$\xspace}

 \defcitealias{cchp2proposal}{CCHP}
 \defcitealias{bea16}{Paper I}
 \defcitealias{hatt17}{Paper II}

 \shorttitle{The TRGB Distance to \ngc}
 \shortauthors{Jang et al.}

\begin{document} 

\title{\textit{The Carnegie-Chicago Hubble Program.} III. THE DISTANCE TO \ngc via the Tip of the Red Giant Branch\footnote{Based in part on observations made with the NASA/ESA \emph{Hubble Space Telescope}, obtained at the Space Telescope Science Institute, which is operated by the Association of Universities for Research in Astronomy, Inc., under NASA contract NAS 5-26555. These observations are associated with program \#13691.}}

\correspondingauthor{In Sung~Jang}\email{isjang@aip.de}

\author[0000-0002-2502-0070]{In Sung~Jang}\affil{Leibniz-Institut f\"{u}r Astrophysik Potsdam, D-14482 Potsdam, Germany}
\affil{Department of Physics \& Astronomy, Seoul National University, Gwanak-gu, Seoul 151-742, Korea}

\author{Dylan~Hatt}\affil{Department of Astronomy \& Astrophysics, University of Chicago, 5640 South Ellis Avenue, Chicago, IL 60637}

\author{Rachael~L.~Beaton}\affil{Observatories of the Carnegie Institution for Science 813 Santa Barbara St., Pasadena, CA~91101}


\author{Myung~Gyoon~Lee}\affil{Department of Physics \& Astronomy, Seoul National University, Gwanak-gu, Seoul 151-742, Korea}

\author{Wendy~L.~Freedman}\affil{Department of Astronomy \& Astrophysics, University of Chicago, 5640 South Ellis Avenue, Chicago, IL 60637}

\author{Barry~F.~Madore}\affil{Department of Astronomy \& Astrophysics, University of Chicago, 5640 South Ellis Avenue, Chicago, IL 60637}\affil{Observatories of the Carnegie Institution for Science 813 Santa Barbara St., Pasadena, CA~91101}

\author{Taylor~J.~Hoyt}\affil{Department of Astronomy \& Astrophysics, University of Chicago, 5640 South Ellis Avenue, Chicago, IL 60637}

\author{Andrew~J.~Monson}\affil{Department of Astronomy \& Astrophysics, Pennsylvania State University, 525 Davey Lab, University Park, PA 16802}

\author{Jeffrey~A.~Rich}\affil{Observatories of the Carnegie Institution for Science 813 Santa Barbara St., Pasadena, CA~91101}

\author{Victoria~Scowcroft}\affil{Department of Physics, University of Bath, Claverton Down, Bath, BA2 7AY, United Kingdom}

\author{Mark~Seibert}\affil{Observatories of the Carnegie Institution for Science 813 Santa Barbara St., Pasadena, CA~91101}

\begin{abstract} 
The Carnegie-Chicago Hubble Program seeks to anchor the distance scale of Type Ia supernovae via the Tip of the Red Giant Branch (TRGB).
Based on deep \emph{Hubble Space Telescope} ACS/WFC imaging, we present an analysis of the TRGB for the metal-poor halo of \ngc, a giant spiral galaxy in the Fornax Cluster that is host to the supernova SN~2012fr.
We have measured its extinction-corrected TRGB magnitude to be \trgbredcorrvalwerr. In advance of future direct calibration by \emph{Gaia}, we set a provisional TRGB luminosity via the Large Magellanic Cloud and find a true distance modulus \truetrgbdmodwerr or \truetrgbdmodMpcwerr. This high-fidelity measurement shows excellent agreement with recent Cepheid-based distances to \ngc and suggests no significant difference in the distances derived from stars of Population~I and II. We revisit the error budget for the \cchp path to the Hubble Constant based on this analysis of one of our most distant hosts, finding a 2.5\% measurement is feasible with our current sample.
\end{abstract} 



\section{Introduction} 

The aim of the Carnegie-Chicago Hubble Program (\cchp) is a direct route to \ho using Type Ia supernovae (\sne)  calibrated entirely via Population (Pop) II stars. The \sne zero point is determined using a distance ladder built from RR Lyrae (RRL) and the Tip of the Red Giant Branch (TRGB) distances to Local Group galaxies. This zero point is then applied to the full sample of \sne in the smooth Hubble flow to arrive at a local, direct estimate of \ho. Eventually, the TRGB will be calibrated in the Galaxy based on \emph{Gaia} trigonometric parallaxes for a three step route to the Hubble constant. 


Since this path is independent of the traditional Pop I Cepheid distance scale that currently sets the \sne zero point, it has the potential to provide insight into the growing (now \textgreater3-$\sigma$) difference in the value of \ho as determined by direct \citep[the distance ladder; e.g.][]{fre12,rie16} and indirect methods \citep[via modeling of the Cosmic Microwave Background; e.g.][]{kom11,planck16}.

Cepheids have long been in use as primary distance indicators: understanding their systematics remains a critical goal. Current uncertainties include the metallicity dependence of the Leavitt Law, the impact of crowding on mean magnitudes, and how best to measure and remove the effects of interstellar extinction \citep[Paper I]{bea16}.
Some, but not all, of these problems relate to the physical location of Cepheids as Pop I stars within the spiral arms of their parent galaxy.
By moving to distance indicators based on Pop II stars, the \cchp aims to bypass these issues of the traditional Cepheid-based extragalactic distance scale by using the intrinsically low-density, metal-poor, and low-extinction regions of galaxies. 

In \citetalias{bea16}, the motivations and full scope of the \cchp were explained in detail.
Taking into account current and projected calibrations of the Pop II distance scale, \citetalias{bea16} estimated a 2.9\% measurement of the Hubble Constant was feasible at the conclusion of the \cchp assuming $\sim$0.1 mag precision of TRGB-based distance measurements in \sne host galaxies.
With direct calibration of the TRGB with \emph{Gaia}, the end precision in the Hubble Constant will be 2.3\% (still assuming 0.1 mag precision of the TRGB). 

In \citet[][Paper II]{hatt17}, the methods for image processing, photometry, and measuring the TRGB for \cchp targets were described in detail and applied to the nearby dwarf irregular galaxy, IC\,1613. In that study, both the TRGB and RRL were used in concert to measure the distance to the galaxy to precisions of 2.3\% and 3.2\%, respectively. Due to its low surface density and proximity to the Galaxy, IC\,1613 represents an ideal case for the application of the tools used in the \cchp. 

It is the purpose of this paper to apply the \citetalias{hatt17} TRGB methodology to one of the most distant \sne-host galaxies in the \cchp sample, \ngc. \ngc is the brightest spiral galaxy in the Fornax Cluster. Cepheids in this galaxy were discovered for the first time \citep{sil99,mad99} as part of the \emph{Hubble Space Telescope} (\hst) $H_0$ Key Project.  The distance to \ngc derived from Cepheids was adopted as the distance to the Fornax Cluster wherein both the fundamental plane and Tully-Fisher relationships were calibrated \citep{mad98,fre01}. 
 
In 2012, \ngc became even more important for the extragalactic distance scale with the discovery of a \sne, SN~2012fr \citep{klotz_2012}. 
SN~2012fr was discovered and classified sufficiently early to receive extensive follow-up with 594 photometric and 144 spectroscopic data points included in the Open Supernova Database\footnote{Data are available at: \url{https://sne.space/sne/SN2012fr/}} \citep{gui16}.
SN~2012fr was also extensively monitored by the optical+NIR Carnegie Supernova Project (CSP), with a detailed analysis of SN~2012fr to be presented by Contreras et al. (in prep). 
A high-fidelity distance to \ngc is therefore a key component of the calibration of the extragalactic distance scale as defined by multiple techniques. 

The structure of the paper is as follows.
In Section \ref{sec:data} the observations and data processing are described. 
In Section \ref{sec:TRGB}, we describe the detection of the TRGB in \ngc, estimate the uncertainties in our measurement, and determine the distance to \ngc by adopting a provisional TRGB luminosity.
In Section \ref{sec:discussion}, we compare our TRGB methodology to other approaches, compare our distance to those derived from Cepheids, and discuss the implications of our measurement in the context of the goals of the \cchp. The primary results of this work are summarized in Section \ref{sec:conclusion}. Detailed comparisons of the methods used in this paper as compared to similar works are given in the Appendix.

\begin{figure*} 
\centering
\includegraphics[width=0.8\textwidth]{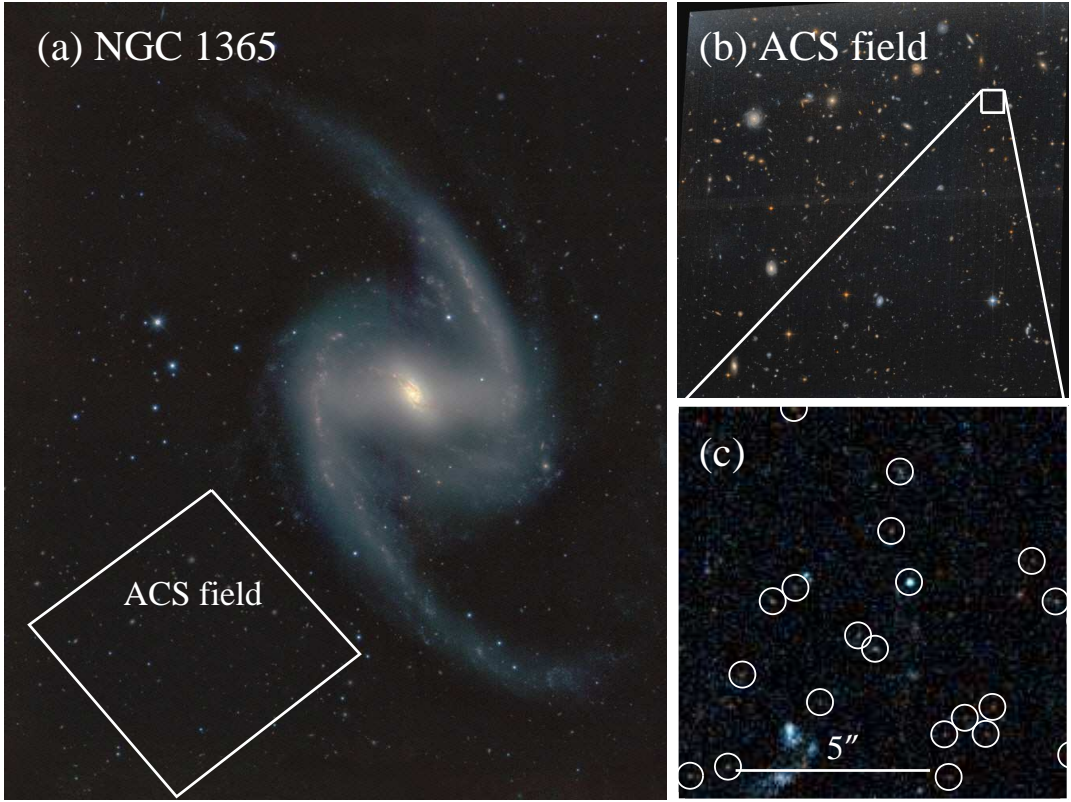}
\caption{
(a) Location of the \hst ACS/WFC field overlaid on a $JHK$ composite image (Contreras et al., in prep.) to demonstrate its position relative to the spiral arms of \ngc.
(b) ACS/WFC F606W and F814W color composite image of the \cchp \ngc field generated with \texttt{DrizzlePac} software. 
(c) A $10\arcsec \times 10\arcsec$ section of the ACS/WFC field. 
Circles enclose point sources, which are primarily red giants in the halo of NGC 1365. Unresolved background galaxies are the primary contaminant for this field. 
}
\label{fig:f1}
\end{figure*} 

\begin{deluxetable*}{ccccccc} 
\tabletypesize{\small} 
\tablewidth{0pt} 
\tablecaption{Summary of ACS/WFC Observations for NGC 1365 \label{tbl:obs_sum}} 
\tablehead{ 
\colhead{Dates} &
\colhead{Filter} &
\colhead{No. obs} &
\colhead{$\alpha$} &
\colhead{$\delta$} &
\colhead{Field Size} &
\colhead{Time (s)}}
\startdata 
2014-09-17 & F606W & 12 & $03^h 33^m 51.4^s$ & $-36^\circ 12\arcmin 05.0\arcsec$ & $3.37\arcmin\times 3.37\arcmin\xspace$ & $\sim1200$ \\
2014-09-21 & F814W & 10 & \ldots & \ldots & \ldots & \ldots \\
2014-09-25 & F814W & 10 & \ldots & \ldots & \ldots & \ldots \\
\enddata 
\tablecomments{See also Figure \ref{fig:f1} for imaging coverage.} 
\end{deluxetable*} 

\section{Data}\label{sec:data} 

The image processing and photometry are performed identically to that described by \citetalias{hatt17} and will be summarized in the subsections to follow. 
A detailed description of the image analysis and photometry pipeline will be presented in a forthcoming work (Beaton et al.~in prep). 

\subsection{Observations and Image Preparation}\label{ssec:obs} 

We obtained optical imaging over 16 orbits on 2014 September 17, 21, and 25 using the ACS/WFC instrument aboard \hst \citep[PID:GO13691, PI: Freedman;][]{cchp2proposal}. Six and ten orbits were used for the F606W and F814W filters, respectively. Pointings were centered on $\mathrm{RA}=3^h 33^m 52.4^s$ and $\mathrm{Dec}=-36^h 12^m 05.0^s$, which is $5\farcm0$ southeast of the NGC 1365 center. The field was selected to be safely in the stellar halo of \ngc and care was taken to place the pointing sufficiently far from the spiral arm by inspection of \emph{WISE} and \emph{GALEX} imaging as described in \citetalias{bea16}. 
Figure \ref{fig:f1}a shows the ACS/WFC pointing relative to the galaxy using a wide-area ($11' \times 11'$) $JHK$ composite image from the FourStar NIR-imager on the Magellan-Baade telescope taken as part of the CSP follow-up campaign for SN~2012fr \citep[Contreras et al.~in prep; for a description of the instrument see][]{persson_2013}. 
Figure \ref{fig:f1}b is a color image of the ACS/WFC observations based on a `drizzled' co-add, and Figure \ref{fig:f1}c is a $10\arcsec \times 10\arcsec$ region of the ACS/WFC image where individual RGB stars are circled.
Figure \ref{fig:f1}c illustrates that the RGB stars in our halo pointing are well isolated from neighboring sources. 

Exposure times were designed to have a signal-to-noise ratio of 10 in F814W at the anticipated apparent magnitude of the TRGB predicted by previous distances estimates to \ngc \citepalias[see Section 4.2.1 of][]{bea16}.
The F606W signal-to-noise is lower (typically by a factor of 3), but the color is only used to remove contaminants and the lower quality does not strongly affect the TRGB itself. 
This strategy provides reliable photometry to a depth of at least one magnitude below the anticipated TRGB to meet sampling requirements for robust TRGB identification as defined in \citet{mad95}.
Individual exposures were $\sim1200$~sec each for total exposure times 14,676~sec and 24,396~sec for F606W and F814W, respectively. 
A summary of these observations, split by the three \emph{HST} visits, are given in Table \ref{tbl:obs_sum}.

Individual ACS/WFC images were obtained through the \emph{Mikulski Archive for Space Telescopes} archive. 
We use the FLC data type, which are calibrated, flat-fielded, and CTE-corrected in the STScI \texttt{CALACS} pipeline. 
The non-uniform pixel area due to ACS/WFC geometric distortions was corrected using the STScI provided Pixel Area Maps\footnote{\url{http://www.stsci.edu/hst/acs/analysis/PAMS}}. 
All further analysis is conducted on these pixel-area corrected FLC frames.

\subsection{Photometry} \label{sec:phot} 

Instrumental magnitudes were derived for individual FLC images via point-spread-function (PSF) fitting in the \textsc{DAOPHOT} software \citep{1987PASP...99..191S}.
We used \textsc{DAOPHOT} to model the PSF for F606W and F814W on synthetic Tiny Tim based star grids (a detailed description of this process will be given in Beaton et al. in prep.).
A direct test of the Tiny Tim PSFs against direct frame-by-frame PSF modeling with isolated, bright stars is described in \citetalias{hatt17} and was found to agree within the photometric uncertainties. 
Images were aligned using \textsc{DAOMATCH}/ \textsc{DAOMASTER} that operate on preliminary catalogs \citep{1987PASP...99..191S}. 
We then use a co-add of all images to determine a `master source list' that is used to simultaneous photometer each individual frame using the \textsc{ALLFRAME} software \citep{1994PASP..106..250S}.
This latter procedure was established for increasing the depth of individual frame photometry in the Key Project \citep{fre01,allframecookbook}.

\subsection{Calibration of HST photometry} \label{sssec:hstcal} 

We transformed the instrumental magnitudes to the ACS Vega magnitudes following equations 2 and 4 of \cite{2005PASP..117.1049S}.
A correction from the PSF magnitudes to the 0\farcs5 aperture magnitudes for each CCD chip was determined by comparing the curve of growth generated from aperture magnitudes to the PSF magnitude (also measured at a 0\farcs5 radius).
We find aperture corrections of --0.044 (chip 1) and --0.058 (chip 2) mag for F814W and --0.037 (chip 1) and --0.047 (chip 2) mag for F606W.
We used photometric zero-point values 26.412 mag for F606W and 25.524 mag for F814W, which were provided for a given observation date by the online STScI ACS Zeropoints Calculator\footnote{\url{https://acszero-points.stsci.edu/}}. The 0\farcs5 to infinite aperture correction values are 0.095 mag for F606W and 0.098 mag for F814W \citep{boh16}.

Intensity mean magnitudes for each filter were computed from the individual frame magnitudes with a median based $\sigma$-clip algorithm setting the clip at 2-$\sigma$. 
We additionally apply an image-quality cut using the `sharpness' parameter to isolate stellar sources using the average value determined from the individual image \textsc{ALLFRAME} photometry. 

\begin{figure} 
\centering
\includegraphics[width=\columnwidth]{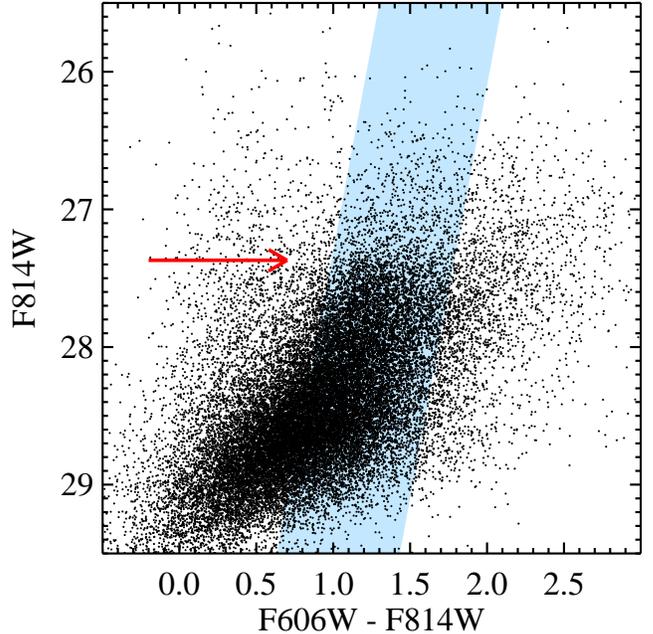}
\caption{CMD of resolved stars in the HST/ACS field of \ngc. An arrow represents the approximate position of the TRGB and the blue shaded region indicates the color range adopted for the red giant branch locus.}

\label{fig:f2}
\end{figure} 

\subsection{Color-Magnitude Diagram} 

The final color-magnitude diagram (CMD) is shown in Figure \ref{fig:f2}.
A change in the source density between $27.3 \lesssim \mathrm{F814W} \lesssim 27.4 $ mag, corresponding to the TRGB, is visible to the eye and highlighted by an arrow. The stars brighter than the TRGB stars are likely thermally pulsing asymptotic giant branch (TP-AGB) stars or blended RGB stars.
We perform an additional step of source filtering by visually inspecting the individual sources in the CMD $\pm$0.5 mag around the TRGB and remove $\sim150$ spurious sources that were components of background galaxies or fringes of bright stars, or $\sim2\%$ of the sources within this magnitude range.

\begin{figure} 
\centering
\includegraphics[width=\columnwidth]{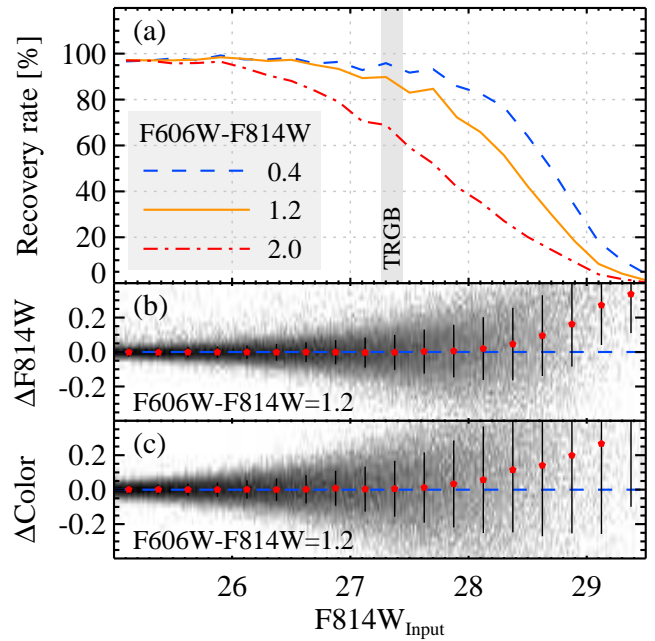} 
\caption{The reliability of the \ngc photometric catalogs. 
(a) Recovery rate versus F814W magnitude for F606W--F814W = 0.4 (dashed line), 1.2 (solid line), and 2.0 (dot-dashed line) derived from artificial star experiments. (b) Difference between input and output F814W magnitudes (input minus output) versus input F814W magnitude. Circles with error bars represent the mean values. (c) Same as in (b) but for the F606W--F814W color. 
A vertical shaded region in each panel indicates the TRGB level of NGC 1365.
}
\label{fig:f3}
\end{figure}

To determine the reliability and completeness of our photometry, we used extensive artificial star tests spanning a range of (F606W-F814W) colors and across the full magnitude range of the CMD in Figure \ref{fig:f2}. 
We input stars with uniform sampling over the range of 25 mag \textless F814W \textless 30 mag and having (F606W-F814W) colors of 0.4, 1.2 and 2.0 mag that span the range of the RGB in our data.
We perform photometry in an identical manner to that described previously and compare the input and output star magnitudes.
Figure \ref{fig:f3}a shows the completeness of the artificial RGB stars as a function of F814W magnitude for F606W-F814W colors of 0.4, 1.2, and 2.0 mag. As anticipated by the F606W signal-to-noise, the photometry is less complete for redder stars, but for $0.4\lesssim \mathrm{F606W}-\mathrm{F814W} \lesssim 1.2$~mag the photometry is 80\% complete at an input magnitude of F814W = 28 mag, well fainter than the visually identified TRGB in Figure \ref{fig:f2}. In Figures \ref{fig:f3}b and \ref{fig:f3}c the recovered photometry is compared to the input photometry for $\mathrm{F606W}-\mathrm{F814W}=1.2$~mag for the F814W magnitude and the $\mathrm{F606W}-\mathrm{F814W}$ color, respectively. 
We find that to input magnitude of $\mathrm{F814W} = 28$~mag, the recovered photometry and colors are in strong agreement. We note that the completeness of our artificial stars drops below $\sim70\%$ for $\mathrm{F814W}\gtrsim28.0$~mag. 

We estimate that the slope of the RGB branch in the \hst flight magnitude system is -6 mag color$^{-1}$, which is steeper than the slope of the $VI$ Johnson-Cousins RGB found in \citetalias{hatt17} from ground-based imaging, but is similar to the slope derived from \cchp fields in M\,31 in the same photometric system (Hatt et al.~in prep.).
We also note that there is noticeable contamination of the measured RGB in Figure \ref{fig:f2}, which makes it difficult to determine the width of the RGB empirically. We manually adjust a color-magnitude cut until we have visually maximized the number of RGB stars encompassed by the cut. The resulting region is shaded in Figure \ref{fig:f2}. The measured RGB, bounded by this color-magnitude range, consists of $\sim$$4,300$ stars, well above the minimum RGB population limits discussed in \citet{mad95} for a robust detection of the TRGB. In the next section, we make our measurement of the TRGB and estimate its associated uncertainties.

\section{The Tip of the Red Giant Branch}\label{sec:TRGB} 

We now estimate a distance to NGC 1365 based on the TRGB method. 
The TRGB is the discontinuity in the RGB luminosity function (LF) caused by the sudden lifting of degeneracy in the He-burning cores of RGB stars \citep[a theoretical overview of RGB evolution can be found in][]{iben_1984,salaris_1997}. The sequence of stars ascending the RGB during core He-burning are thus truncated at this magnitude as they rapidly evolve away from the RGB sequence. 
As first shown empirically for a sample of nearby galaxies by \citet{lee93}, the TRGB is well-delineated and effectively flat for metal-poor populations in the $I$-band, which is equivalent to the F814W filter in the \hst flight magnitude system. 

The algorithmic approach to measuring the TRGB has been refined and expanded since its initial implementation in \cite{lee93}. A review of published techniques since that time is given in  \citetalias{hatt17}. In this study, we follow the method outlined in  \citetalias{hatt17}. 
The general approach to our TRGB measurement is as follows: 
First, the RGB LF is binned in 0.01 F814W mag bins, where we have isolated stars using color-magnitude and image-quality cuts (Figure \ref{fig:f2}). 
The finely binned LF is then smoothed using GLOESS (Gaussian-windowed, Locally-Weighted Scatterplot Smoothing), which is a data smoothing technique first introduced in an astrophysical context by \citet{per04} for Cepheid light curves and described in more detail in \citet{monson_2017} for RR Lyrae light curves. 
The technique uses a smoothing window around a reference point in the input discrete function and applies a Gaussian weighting function based on the distance to neighboring data points, which is set by a scaling parameter, $\sigma_{s}$. 
This smoothed LF is then convolved with an edge detection kernel and, as in \citetalias{hatt17}, we use the Sobel filter, $[-1, 0, +1]$, which is derived from finite-difference methods and is a simple approximation to the first-derivative of a discrete function. The edge detector will produce the largest response when the change in the LF is the greatest, i.e. at the discontinuity present at the TRGB. 

As discussed \citetalias{hatt17}, there are practical considerations for application of this technique that must be statistically modeled for a given dataset. 
The primary concern is the selection of an optimal size for the Gaussian scaling parameter $\sigma_{s}$, which is determined as the value for which the combination of the statistical and systematic uncertainties associated with the TRGB edge-detection are minimized.
\citetalias{hatt17} describes a procedure using artificial star tests to empirically derive $\sigma_{s}$ and the associated uncertainties, which we apply to \ngc in Section \ref{sec:trgb_optimize}. 
In Section \ref{sssec:trgb_meas} we then measure the \ngc TRGB and determine our final distance in Section \ref{sec:distance}.

\subsection{Optimizing the TRGB Detection} \label{sec:trgb_optimize} 

In order to make a robust measurement of the TRGB, we seek the optimal leveling of smoothing in the LF that  reduces the statistical and systematic errors. Sections \ref{sssec:aslf} and \ref{sssec:edge_sim} describe the creation of an artificial star luminosity function (ASLF) and simulations to model the properties of our GLOESS smoothing function and the [-1,0,+1] kernel.

\subsubsection{Artificial Stars and Luminosity Functions}\label{sssec:aslf} 

We created an artificial star luminosity function (ASLF) to estimate the systematic bias and completeness of our photometry. We assumed that the luminosity function (LF) for the RGB has a of slope $0.3\pm0.04$ dex mag$^{-1}$ \citep[see][]{men02}. Our ASLF begins at the estimated tip magnitude $\mathrm{F814W}\approx19.33$~mag in the instrumental magnitude system, or $\mathrm{F814W}=27.36$ in the ACS Vega system, and it extends to $\mathrm{F814W}\approx20.33$~mag or $\mathrm{F814W}=28.36$~mag in the instrumental and ACS Vega systems, respectively. We assign a fixed color of $\mathrm{F814W}-\mathrm{F606W}$ = 1. One-thousand stars were sampled at random from this ASLF distribution and placed into each individual FLC frame at pixel coordinates uniformly distributed in $X$ and $Y$. These stars were manually added to the `master list' of sources and the \textsc{ALLFRAME} photometry was performed as previously described. The artificial star process was repeated 50 times, producing a total of $50,000$ artificial RGB stars for which $\sim$$42,500$ were successfully measured (85\% completeness over the RGB magnitude range). Figure \ref{fig:art_stars}a shows the input and output ASLFs as yellow and blue histograms, respectively. While the input ASLF has a hard bright edge to represent the TRGB, the output ASLF illustrates both incompleteness across the LF and broadening of the TRGB due to measurement uncertainties.

\subsubsection{Simulating TRGB edge detections}\label{sssec:edge_sim} 

\begin{figure} 
\centering
\includegraphics[width=\columnwidth]{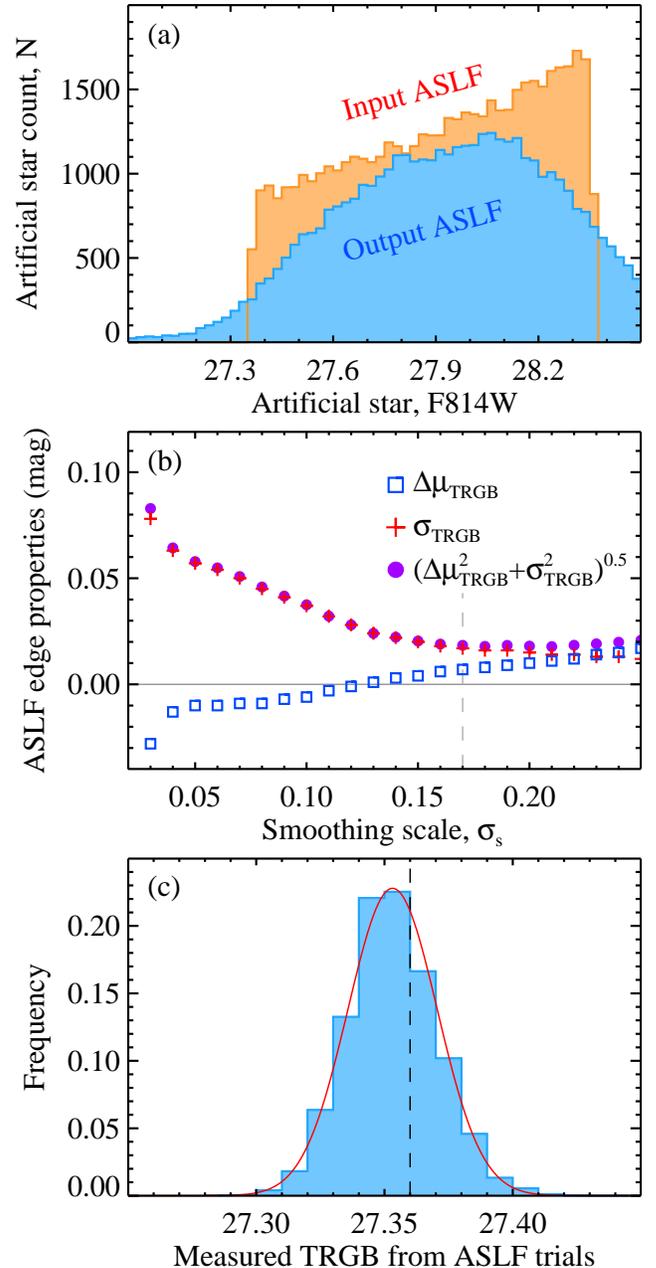}
\caption{\label{fig:art_stars}
 Artificial star tests are used to determine to determine the optimal LF smoothing scale as well as derive the statistical and systematic uncertainties in our TRGB measurement.
(a) The input (orange) and recovered (blue) artificial luminosity functions.
(b) Displacement of the input TRGB ($\mu_{\mathrm{TRGB}}$, open blue squares), the dispersion in measured TRGB ($\sigma_{\mathrm{TRGB}}$, red plus symbols), and the quadrature sum of these values (purple filled circles). The systematic and random (statistical) uncertainties in our edge-detection are represented by $\mu_{\mathrm{TRGB}}$ and $\sigma_{\mathrm{TRGB}}$, respectively. 
The optimal \sigmasmooth yields minimum total uncertainty and is marked by a vertical dashed line.
(c) Distribution of maximal Sobel kernel responses for our 10,000 realizations of the ASLF for the optimal smoothing scale of $\sigma_s=0.17$ mag (blue histogram) with the Gaussian model of the distribution over plotted (red line). A vertical dashed line marks the input TRGB magnitude.}

\end{figure}

We now quantify the statistical and systematics errors associated with GLOESS and our $[-1,0,+1]$ edge detection kernel. We restrict our ASLF to the color-magnitude constraints visualized in Figure \ref{fig:f2} (blue box), as described in the previous section, and randomly select $4,300$ stars with replacement from this sample to simulate the sample size defining the TRGB in our \ngc data. We construct a LF using 0.01 mag bins and apply GLOESS with a fixed value for \sigmasmooth. We apply the Sobel kernel on the smoothed LF and we select the bin of greatest response as the TRGB. We repeat this process 10,000 times each for 0.01 \textless~\sigmasmooth~\textless~0.25~mag in 0.01 mag increments. We use the distribution of TRGB measurements to estimate the intrinsic uncertainties of the GLOESS smoothing and Sobel kernel. The displacement of the detected edge, $\mu_{\mathrm{TRGB}}$, for a given \sigmasmooth is defined as the mean offset from the TRGB edge (Figure \ref{fig:art_stars}a) and serves as an estimate of the systematic uncertainty for a given \sigmasmooth. The dispersion of estimates, $\sigma_{\mathrm{TRGB}}$, is the $\pm1\sigma$ standard deviation of all realizations and serves as our estimate of the random (statistical) uncertainty for the \sigmasmooth.


Figure \ref{fig:art_stars}b gives the results for all \sigmasmooth. At \sigmasmooth$\approx0.17$ mag the combined error (the quadrature sum of $\mu_{\mathrm{TRGB}}$ and $\sigma_{\mathrm{TRGB}}$) is minimized and this represents the `optimal' smoothing scale (e.g., the scale that yields the smallest total uncertainty). Figure \ref{fig:art_stars}c shows the distribution of measured TRGB value for this \sigmasmooth. To measure the uncertainties with this smoothing scale, we fit a Gaussian to the resulting distribution of TRGB measurements and adopt the offset from the input and the width as the systematic and random uncertainties \trgbobsvalsyserr~ and \trgbobsvalstaterr, respectively.

These errors associated with the measurement of the idealized \ngc LF are remarkably small. In \citetalias{hatt17}, it was shown that the TRGB magnitude for IC\,1613 could be constrained to only $\approx0.02$~mag. If the uncertainties were based solely on the photometric errors for its TRGB stars, one might expect that the \ngc TRGB measurement would have larger uncertainties since IC\,1613 lies at $\sim730$ kpc compared to the anticipated \ngc distance $\sim 18$ Mpc. However, although the overall uncertainty in the TRGB measurement for IC\,1613 is comparable to the photometric errors of its individual TRGB stars, the number of stars defining TRGB also plays a large role in its detectability: the greater the sample of stars contributing to the tip, the more readily it is detected.  The \ngc RGB in this study is over three times more populated than IC\,1613 in \citetalias{hatt17}. We are undertaking a series of simulations to further explore and quantify these issues, to be published in future.


\subsection{Measurement of the TRGB}\label{sssec:trgb_meas} 

\begin{figure*} 
\centering
\includegraphics[angle=0,width=0.7\textwidth]{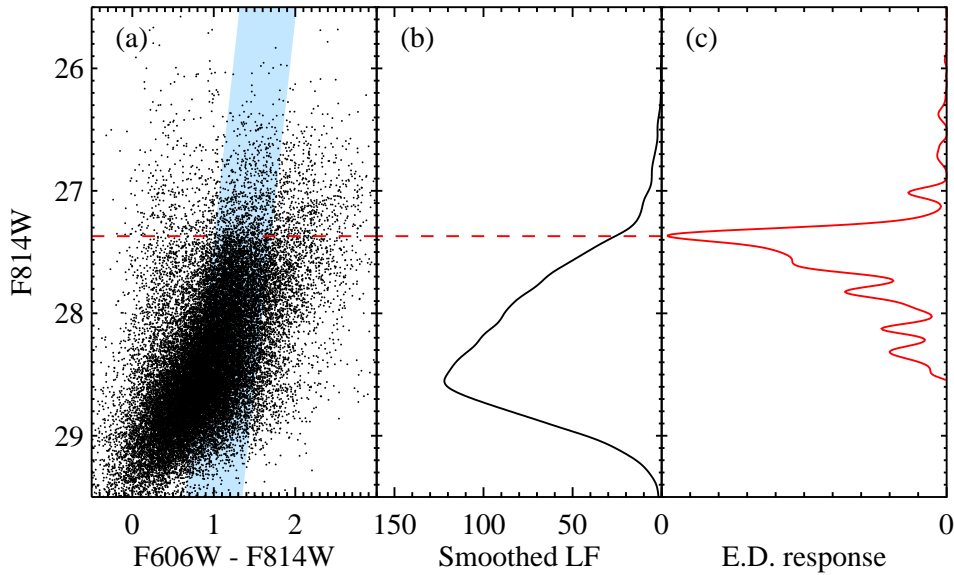}
\caption{\ngc TRGB edge detection: (a) CMD of sources in \ngc after filtering based on image quality. The blue shaded region indicates the selected parameter space that best encompasses the RGB and limits the TRGB to where the F814W tip magnitude is flat with color. 
(b) GLOESS smoothed luminosity function in the F814W band using the optimal smoothing parameter determined via ASLF simulations; and  
(c) Edge response of the $[-1,0,+1]$ kernel with signal-to-noise weighting find a peak response at $\mathrm{F814W}=$ \trgbobsvalrounded ~mag, which is indicated in panels (a) and (b) by the dashed line.
\label{fig:n1365trgbmeas}}
\end{figure*}

Figure \ref{fig:n1365trgbmeas}a presents the final CMD used to determine the distance to \ngc. We apply the color-magnitude restrictions, described in the previous sections, that isolate the RGB and are indicated by the blue shading in Figure \ref{fig:n1365trgbmeas}a. These limits coincide with the color range over which the TRGB magnitude is known to be flat with color \citep{lee93, jan17a}. 

Figure \ref{fig:n1365trgbmeas}b is the resulting LF for stars in the blue shaded region after smoothing using the GLOESS algorithm and our optimal scaling parameter, $\sigma_{s} =$ 0.17 mag, as determined in the previous section. Figure \ref{fig:n1365trgbmeas}c is the result of applying the [-1,0,+1] Sobel kernel to the LF, which shows a strong peak at \trgbobsval ~mag (indicated by the dashed lines in Figures \ref{fig:n1365trgbmeas}a and \ref{fig:n1365trgbmeas}b). Based on the simulations in the previous subsection, we assign a statistical uncertainty of \trgbobsvalstaterr ~mag and a systematic uncertainty of \trgbobsvalsyserr ~mag. Our final TRGB determination is \trgbobsvalwerr, before correcting for line-of-sight reddening.

\subsection{TRGB Reddening and Distance} \label{sec:distance} 

The Milky Way foreground extinction is estimated to be small: \Vextinction~mag for F606W and \Iextinction~mag for F814W, or $E(\mathrm{F606W}-\mathrm{F814W})$ = 0.020~mag \citep[][retrieved from NED]{sch11}. Applying these estimates to our TRGB measurement from the previous subsection, we find an extincted-corrected TRGB magnitude of F814W~=~\trgbredcorrval~mag. 

Currently, the absolute magnitude of the TRGB has no direct trigonometric calibration, though \emph{Gaia} parallaxes will provide one in the near future. In the interim, we have chosen to adopt an absolute magnitude for the TRGB for the \cchp analyses. The derivation of this value will be presented in a forthcoming analysis of  the main body of the Large Magellanic Cloud (LMC) and is anchored to both Cepheids and eclipsing binaries (Freedman et al.~in prep). The zero point is \trgblumwerr and was used in \citetalias{hatt17} for the \cchp distance to IC\,1613. The adoption of a provisional zero-point is a strategy similar to that employed in the mid-stages of the Key Project for the purpose of internal consistency. This provisional calibration is also consistent to within $\pm$1-$\sigma$ with the canonical TRGB calibration based on globular clusters, $M_{I}^\mathrm{TRGB}\approx-4$~mag, and has held up in more detailed calibration efforts \citep[e.g.,][among others]{riz07,jan17a}. We note that \citet{jan17a} find $M_{I}^\mathrm{TRGB}$ = --3.970 $\pm$ 0.102 mag in the LMC using a similar process (e.g., spanning a similar color range).

Applying this provisional zero-point to our TRGB apparent magnitude, we find a true distance modulus to \ngc of \truetrgbdmodwerr, or a distance of \truetrgbdmodMpcwerr. 
Table \ref{tab_distance} summarizes the values for the TRGB magnitude, the distance modulus,  its uncertainties, and the adopted reddening value.

\begin{deluxetable}{lccc} 
\tabletypesize{\scriptsize}
\setlength{\tabcolsep}{0.05in}
\tablecaption{TRGB Distance and Error Budget \label{tab_distance}}
\tablewidth{0pt}
\tablehead{ \colhead{Parameter}  &  \colhead{Value} & \colhead{$\sigma_{ran}$} & \colhead{$\sigma_{sys}$} }
\startdata
TRGB F814W magnitude 	          & \trgbobsvalrounded  & \trgbobsvalstaterr & \trgbobsvalsyserr \\
$A_{\mathrm{F814W}}$	          & \Iextinction        & \nodata            & \nodata \\
Provisional $M_I^{\mathrm{TRGB}}$ & \trgblum            & \trgblumstaterr    & \trgblumsyserr \\
\hline
True distance modulus [mag]       & \truetrgbdmod & \dmodcombinedstaterr & \dmodcombinedsyserr\\
{\bf Distance} [Mpc]              & \truetrgbdmodMpc & \truetrgbdmodMpcstaterr & \truetrgbdmodMpcsyserr\\
\hline \hline
\enddata
\end{deluxetable} 

\section{Discussion} \label{sec:discussion} 

In this section we provide context for our TRGB measurement with regard to existing Cepheid-based distances and the goals of the \cchp.
First, we compare the \cchp methods to recent TRGB studies at a similar distance to \ngc in Section \ref{ssec:trgbcomp}.
Next, we compare the \cchp TRGB distance to those determined by Cepheids in Section \ref{ssec:distcomp}.
Lastly, in Section \ref{ssec:cchpgoals} we discuss the how the results of this study impact the goals of the \cchp. 

\subsection{Comparison to Other TRGB Studies} \label{ssec:trgbcomp} 

The objective of the CCHP is the measure of the Hubble constant to high fidelity, minimizing systematics by observing and applying a homogeneous analysis of the TRGB in galaxies spanning 10 magnitudes in distance modulus \citepalias[see][Table 5 for a summary of the TRGB targets]{bea16}. We have developed a data-reduction strategy that can be applied to galaxies spanning this wide range in distance. 
As a result, the data processing, treatment of the sloped TRGB, and edge-detection strategies differ from similar studies using the TRGB at these distances. 
In the subsections to follow, we compare our methods to those used in other studies. 

\subsubsection{Data Processing}\label{ssec:technique} 

Previous studies using the TRGB method at the \ngc distance \citep[e.g.,][among others]{jan17a,jan17b} have utilized stacks produced by the STScI \texttt{DrizzlePac} software \citep{drizzle2015}, from which photometry is derived using point-spread function fitting to bright stellar sources in the image.
These stacks provide image products that can be optimized in resolution and provide higher signal-to-noise than analyses performed on individual frames, but come at the cost of producing image products that vary based on the observing strategy employed. 

We provide a detailed comparison to photometry derived identically to \citet{jan17b} in Appendix \ref{app:redux}. We find our photometry to be statistically identical over the magnitude range of interest. Moreover, the same TRGB magnitude is obtained within the statistical uncertainties. 
Thus, we find no bias due to our reduction strategy. 

\subsubsection{Rectification of Sloped TRGB}\label{ssec:retify} 

Recent studies applying the TRGB at a similar distance to \ngc \citep[e.g.,][]{jan17a,jan17b} have employed a technique that allows the higher-metallicity stars, for which the tip magnitude is fainter as a function of metallicity, to be used in the TRGB detection and thereby have better statistics at the tip magnitude. The form of this correction is a normalization of the tip magnitude as a function of color that effectively rectifies the slanted or curving, metal-rich portion of the TRGB into a sharp edge. The form of the rectification is either linear \citep{mad09,riz07} or quadratic \citep{jan17a} with color.
Because the \cchp program has specifically designed pointings to target the metal-poor halos of galaxies and the signal-to-noise in our F606W is 1/3 that in the F814W, we opt to not rectify the F814W magnitudes. We do, however, provide a detailed comparison to the application of these methods, and to the body of work summarized in \citet{jan17b}, in Appendix \ref{app:retified}. We find the results using the rectified magnitudes to be identical to our non-rectified magnitudes within the uncertainties, and find no bias due to our choice to limit the color range used in our LF. 

\subsubsection{Edge Detectors} \label{ssec:edge} 

We have followed a simple edge detection methodology for the TRGB in this work, modeled after \citetalias{hatt17}, for the ease of estimating the uncertainties associated with our measurement as well as avoiding previous algorithmic complications such as binning and over-smoothing data. 
As with \citetalias{hatt17}, we compare results using several of the different approaches in Appendix \ref{app:edge}. We find that there is good agreement with the TRGB measurement presented in this study.

\subsubsection{Summary} \label{ssec:sum} 

In this subsection, we compare the methods adopted in this work (and in \citetalias{hatt17}) to those commonly used in the literature, in particular the body of work encompassed by \citet{jan17b}. This comparison was completed in three phases: (i) the image processing and photometry, (ii) correcting the metal-rich slope of the TRGB, and (iii) testing alternate edge detectors. For all tests, we find our methods to agree within the statistical uncertainties and thus conclude that our techniques are both sufficient for the goals of the \cchp and consistent with techniques used by other authors. 

\subsection{Comparison to Cepheid Distances} \label{ssec:distcomp} 

Previously published distance modulus estimates to \ngc   based on Cepheids, Type II supernova (SN 2001du), and the Tully-Fisher relation (NED-D) range from $\mu_0=29.52$ mag to 32.09 mag with a mean and median of 31.20 mag and 31.26 mag, respectively. Cepheids are the only fully independent measure of distance to \ngc, and we therefore focus our distance comparison on them.

There are roughly 30 distance estimates for NGC 1365 based on Cepheids in NED circa 2017, though nearly all of these estimates are based on the same image dataset that was obtained for the Hubble Key Project: 12 epochs of F555W and 4 epochs of F814W taken with the WFPC2 instrument \citep{fre01, sil99}, with later works updating the calibration of the original results. These updated distance moduli show a large range from $\mu_0=31.18$ to 32.09, resulting primarily from uncertainties in the color and metallicity dependency of the Cepheids.  Because of the uncertainty in the Cepheid calibration, and the bias introduced by comparing the results of consecutive publications differing only in zero-points, we have chosen the \citet{fre01} result to represent the results from this ensemble of publications, consistent with the approach of \citetalias{hatt17}. We have further considered a recent analysis by \citet{rie16}, who analyze new NIR photometry for a subset of the Cepheids originally discovered within the Key Project \citep{fre01}.

\begin{figure} 
\centering
\includegraphics[width=\columnwidth]{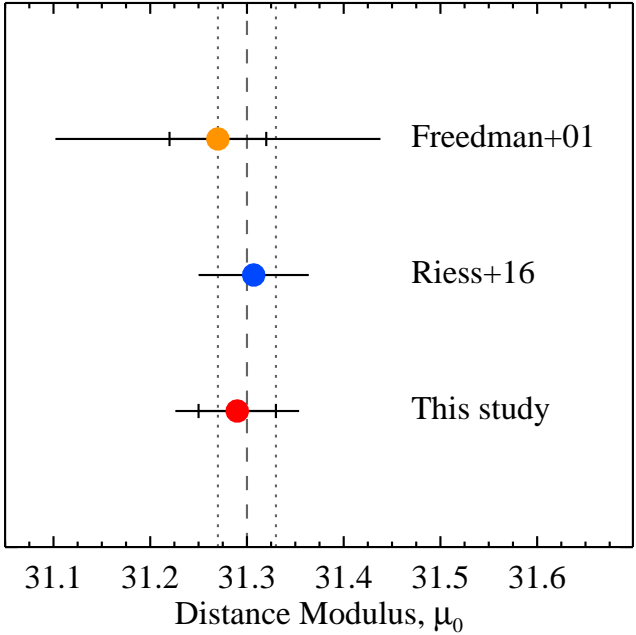}
\caption{Comparison of Cepheid distances to \ngc and the TRGB distance of this study. Vertical dashed line and dotted lines show the weighted average distance and $\pm\,$1-$\sigma$ confidence intervals, respectively, or \wavgdistance~mag. 
The results of independent analyses of \ngc agree remarkably well.}
\label{fig:distance_estimates}
\end{figure}

The final KP distance was $\mu_0$=31.27 $\pm$ 0.05 $\pm$ 0.14 \citep{fre01} and was anchored to the LMC. 
\citet{rie16} find $\mu_0$=31.307 $\pm$ 0.057 mag using NIR Cepheids and anchoring the zero-point of the PL to a number of different techniques in the Galaxy, M\,31, and NGC\,4258\footnote{We refer the reader to that work for the full description of their anchoring process and tests thereof.}. 
Figure \ref{fig:distance_estimates} illustrates the consistency of the two independent Cepheid distances with that derived in this study. The sample error on the mean is only 0.03 mag, and gives no indication of a significant difference in the distances derived from stars of Pop~I and II for \ngc. The weighted-average of these results suggests a true distance modulus \wavgdistance, which is statistically indistinguishable from the TRGB measurement presented here based on the provisional TRGB luminosity in the LMC (Freedman et al.~in prep).

\subsection{Evaluating the \cchp} \label{ssec:cchpgoals} 

\subsubsection{Comparing the Pop I and Pop II Scales} 
A primary goal of the \cchp is to provide a test of the systematics of the Cepheid-based distance calibration for \sne.
In \citetalias{hatt17}, we found consistency between the Pop I (Cepheids) and the Pop II (RRL and TRGB) based on distances to the Local Group dwarf irregular galaxy, IC\,1613. 
The Cepheids in IC\,1613 represent a sample with low crowding, low metallicity \citep[12+$\log$(O/H)=7.90;][]{bresolin_2007}, and low internal reddening.
Thus, in addition to being an ideal case for the TRGB, IC\,1613 is also an ideal galaxy for application of the Leavitt Law.
In contrast, \ngc presents more challenges  for accurately measuring  Cepheids; it has high crowding in the spiral arms, is of solar or super-solar metallicity \citep[8.33\textless12+$\log$(O/H)\textless8.71;][]{bresolin_2005}, and the internal reddening is larger than that of IC\,1613. 
Thus, we can provide an initial assessment on the impact of these effects on the Cepheid distance scale.

As discussed previously, we find broad agreement between the Pop I and Pop II distance indicators for both the simple case of IC\,1613 \citepalias{hatt17} and for the more complicated case of \ngc (this work). 
The LMC has an intermediate metallicity, 12+$\log$(O/H) = 8.26, \citep[estimated using identical techniques to those in \ngc and IC\,1613 by][]{berg_2012}. In the LMC, the Cepheid and (geometric) eclipsing binary distances agree to better than 1\%. This early agreement between Pop I and Pop II scales suggests that the oft-cited concerns regarding Cepheids of crowding, metallicity, and extinction cannot be fully responsible for the current impasse between direct and indirect paths to the Hubble Constant. As described in \citetalias{bea16}, over the course of the \cchp we will provide additional direct tests of Cepheids, RRL, and the TRGB in three Local Group galaxies (with RRL and TRGB tested in an additional three galaxies) and between Cepheids and the TRGB for a total of five \sne hosts.


\subsubsection{The TRGB Error Budget} 
In \citetalias{bea16}, we used literature studies to provide an estimate for the \cchp error budget. 
We adopted a TRGB measurement uncertainty of $\sigma$=0.10 mag, which was justified as twice the uncertainty quoted by \citet{riz07} (to account for increased magnitude uncertainties for our more distant objects) and the uncertainty determined by \citet{caldwell_2006} for a sample of Virgo dwarf galaxies. 
To this we added in quadrature a term for the `blurring' of the TRGB due to multi-metallicity populations of $\sigma_{[Fe/H]}$ = 0.028 mag. 
With results from \ngc and IC\,1613 in hand, we can evaluate these estimates. 

As is demonstrated by our comparison of rectified and non-rectified TRGB magnitudes (Section \ref{ssec:retify}), the metallicity term is likely unnecessary if the color range is sufficiently restricted (as is done here).  
Our total TRGB measurement uncertainty (Table \ref{tab_distance}) is 0.03 mag, a factor of three smaller than that assumed in \citetalias{bea16}. This can be understood largely in the context of the larger sample size populating the TRGB, as was originally described  by \citet{mad95}. With 4300 stars populating the LF below the TRGB, we are able to detect it much more precisely than the \citet{caldwell_2006} study of dwarf galaxies at a similar distance. Moreover, we obtain measurement uncertainties of the same level as \citet{riz07} for their more nearby objects; we are able to cover a larger physical area in the halos of our more distant galaxies and this makes up for the loss in photometric accuracy due to the larger distance. 

If we assume that measurement uncertainties of 0.05 mag can be obtained for each of our nine \sne host galaxies (all of which are no more distant than \ngc), then this is a 50\% reduction in the uncertainty in our initial error budget \citepalias{bea16}. The TRGB uncertainty is added in quadrature to the 0.120 mag intrinsic scatter of the \sne \citep{folatelli_2010} and this results in a total uncertainty for an individual measurement of the \sne absolute magnitude of 0.130 mag and an uncertainty of 0.038 mag (1.73\%) for that term in the averaged zero point for the 12 \sne in our sample. Assuming no other changes to the \cchp error budget (the analyses of \citetalias{hatt17} align with the predictions), this suggests a 2.5\% measure of the Hubble constant pre-\emph{Gaia} in the RRL-TRGB hybrid path will be feasible. A direct calibration of the TRGB skips the RRL rung in the \cchp and the uncertainty in the Hubble constant then approaches the 2\% level. The dominant term in the error budget remains the number of independently calibrated \sne and efforts to expand that number, and in turn provide greater insight into the intrinsic scatter of \sne, will provide the greatest impact on the end precision from this route to the Hubble constant.

\section{Conclusion}\label{sec:conclusion} 

As part of the \cchp, we have measured a TRGB distance to the Fornax Cluster galaxy, \ngc.  We have resolved  old, metal-poor RGB stars in the halo of \ngc with photometry obtained from deep F606W and F814W images taken with the ACS/WFC instrument aboard \hst. We have undertaken an extensive comparison of the different techniques in use for measuring the TRGB, and find that the technique we have adopted for the \cchp is consistent to within the uncertainties.


We have measured an extinction-corrected TRGB \trgbobsvalwerr, which, using a provisional value for the TRGB absolute magnitude, corresponds to a true distance modulus of  \truetrgbdmodwerr or a physical distance of \truetrgbdmodMpcwerr (Table \ref{tab_distance}). 
Our distance estimate is consistent with the existing, independent measurements using the Cepheid Leavitt Law in the optical and near-infrared bands. Taken in the context of similar agreement for IC\,1613 from \citetalias{hatt17}, we find broad agreement between the Pop I and Pop II scales over a large span of Cepheid metallicity, crowding, and internal extinction.

\smallskip

\acknowledgments 
This work was supported by the National Research Foundation of Korea (NRF) grant funded by the Korea Government (MSIP) (No.~2012R1A4A1028713).
Support for program \#13691 was provided by NASA through a grant from the Space Telescope Science Institute, which is operated by the Association of Universities for Research in Astronomy, Inc., under NASA contract NAS 5-26555.
We thank the Carnegie Institution for its continued support of this program over the past 30 years.
This research has made use of the NASA/IPAC Extragalactic Database (NED), which is operated by the Jet Propulsion Laboratory, California Institute of Technology, under contract with the National Aeronautics and Space Administration. 
This paper includes data gathered with the $6.5$\,m Magellan Telescopes located at Las Campanas Observatory, Chile.

\facility{HST (ACS/WFC), Magellan-Baade (FourStar)}
\software{DAOPHOT \citep{1987PASP...99..191S}, ALLFRAME \citep{1994PASP..106..250S}, TinyTim \citep{2011SPIE.8127E..0JK}, DrizzlePac \citep{fru02}}

\bibliographystyle{aasjournal}
\bibliography{ms.bib}

\begin{thebibliography}{}
\expandafter\ifx\csname natexlab\endcsname\relax\def\natexlab#1{#1}\fi

\bibitem[{{Avila} {et~al.}(2015){Avila}, {Hack}, {Cara}, {Borncamp}, {Mack},
  {Smith}, \& {Ubeda}}]{drizzle2015}
{Avila}, R.~J., {Hack}, W., {Cara}, M., {et~al.} 2015, in Astronomical Society
  of the Pacific Conference Series, Vol. 495, Astronomical Data Analysis
  Software an Systems XXIV (ADASS XXIV), ed. A.~R. {Taylor} \& E.~{Rosolowsky},
  281

\bibitem[{{Beaton} {et~al.}(2016){Beaton}, {Freedman}, {Madore}, {Bono},
  {Carlson}, {Clementini}, {Durbin}, {Garofalo}, {Hatt}, {Jang}, {Kollmeier},
  {Lee}, {Monson}, {Rich}, {Scowcroft}, {Seibert}, {Sturch}, \& {Yang}}]{bea16}
{Beaton}, R.~L., {Freedman}, W.~L., {Madore}, B.~F., {et~al.} 2016, \apj, 832,
  210

\bibitem[{{Bellazzini} {et~al.}(2001){Bellazzini}, {Ferraro}, \&
  {Pancino}}]{bel01}
{Bellazzini}, M., {Ferraro}, F.~R., \& {Pancino}, E. 2001, \apj, 556, 635

\bibitem[{{Bellazzini} {et~al.}(2004){Bellazzini}, {Ferraro}, {Sollima},
  {Pancino}, \& {Origlia}}]{bel04}
{Bellazzini}, M., {Ferraro}, F.~R., {Sollima}, A., {Pancino}, E., \& {Origlia},
  L. 2004, \aap, 424, 199

\bibitem[{{Berg} {et~al.}(2012){Berg}, {Skillman}, {Marble}, {van Zee},
  {Engelbracht}, {Lee}, {Kennicutt}, {Calzetti}, {Dale}, \&
  {Johnson}}]{berg_2012}
{Berg}, D.~A., {Skillman}, E.~D., {Marble}, A.~R., {et~al.} 2012, \apj, 754, 98

\bibitem[{{Bohlin}(2016)}]{boh16}
{Bohlin}, R.~C. 2016, \aj, 152, 60

\bibitem[{{Bresolin} {et~al.}(2005){Bresolin}, {Schaerer}, {Gonz{\'a}lez
  Delgado}, \& {Stasi{\'n}ska}}]{bresolin_2005}
{Bresolin}, F., {Schaerer}, D., {Gonz{\'a}lez Delgado}, R.~M., \&
  {Stasi{\'n}ska}, G. 2005, \aap, 441, 981

\bibitem[{{Bresolin} {et~al.}(2007){Bresolin}, {Urbaneja}, {Gieren},
  {Pietrzy{\'n}ski}, \& {Kudritzki}}]{bresolin_2007}
{Bresolin}, F., {Urbaneja}, M.~A., {Gieren}, W., {Pietrzy{\'n}ski}, G., \&
  {Kudritzki}, R.-P. 2007, \apj, 671, 2028

\bibitem[{{Caldwell}(2006)}]{caldwell_2006}
{Caldwell}, N. 2006, \apj, 651, 822

\bibitem[{{Cioni} {et~al.}(2000){Cioni}, {van der Marel}, {Loup}, \&
  {Habing}}]{cio00}
{Cioni}, M.-R.~L., {van der Marel}, R.~P., {Loup}, C., \& {Habing}, H.~J. 2000,
  \aap, 359, 601

\bibitem[{{Conn} {et~al.}(2011){Conn}, {Lewis}, {Ibata}, {Parker}, {Zucker},
  {McConnachie}, {Martin}, {Irwin}, {Tanvir}, {Fardal}, \& {Ferguson}}]{con11}
{Conn}, A.~R., {Lewis}, G.~F., {Ibata}, R.~A., {et~al.} 2011, \apj, 740, 69

\bibitem[{{Folatelli} {et~al.}(2010){Folatelli}, {Phillips}, {Burns},
  {Contreras}, {Hamuy}, {Freedman}, {Persson}, {Stritzinger}, {Suntzeff},
  {Krisciunas}, {Boldt}, {Gonz{\'a}lez}, {Krzeminski}, {Morrell}, {Roth},
  {Salgado}, {Madore}, {Murphy}, {Wyatt}, {Li}, {Filippenko}, \&
  {Miller}}]{folatelli_2010}
{Folatelli}, G., {Phillips}, M.~M., {Burns}, C.~R., {et~al.} 2010, \aj, 139,
  120

\bibitem[{{Frayn} \& {Gilmore}(2003)}]{fra03}
{Frayn}, C.~M., \& {Gilmore}, G.~F. 2003, \mnras, 339, 887

\bibitem[{{Freedman}(2014)}]{cchp2proposal}
{Freedman}, W. 2014, {CHP-II: The Carnegie Hubble Program to Measure Ho to 3\%
  Using Population II}, HST Proposal, ,

\bibitem[{{Freedman} {et~al.}(2012){Freedman}, {Madore}, {Scowcroft}, {Burns},
  {Monson}, {Persson}, {Seibert}, \& {Rigby}}]{fre12}
{Freedman}, W.~L., {Madore}, B.~F., {Scowcroft}, V., {et~al.} 2012, \apj, 758,
  24

\bibitem[{{Freedman} {et~al.}(2001){Freedman}, {Madore}, {Gibson}, {Ferrarese},
  {Kelson}, {Sakai}, {Mould}, {Kennicutt}, {Ford}, {Graham}, {Huchra},
  {Hughes}, {Illingworth}, {Macri}, \& {Stetson}}]{fre01}
{Freedman}, W.~L., {Madore}, B.~F., {Gibson}, B.~K., {et~al.} 2001, \apj, 553,
  47

\bibitem[{{Fruchter} \& {Hook}(2002)}]{fru02}
{Fruchter}, A.~S., \& {Hook}, R.~N. 2002, \pasp, 114, 144

\bibitem[{{Guillochon} {et~al.}(2017){Guillochon}, {Parrent}, {Kelley}, \&
  {Margutti}}]{gui16}
{Guillochon}, J., {Parrent}, J., {Kelley}, L.~Z., \& {Margutti}, R. 2017, \apj,
  835, 64

\bibitem[{{Hatt} {et~al.}(2017){Hatt}, {Beaton}, {Freedman}, {Madore}, {Jang},
  {Hoyt}, {Lee}, {Monson}, {Rich}, {Scowcroft}, \& {Seibert}}]{hatt17}
{Hatt}, D., {Beaton}, R.~L., {Freedman}, W.~L., {et~al.} 2017, ArXiv e-prints,
  arXiv:1703.06468

\bibitem[{{Iben} \& {Renzini}(1984)}]{iben_1984}
{Iben}, I., \& {Renzini}, A. 1984, \physrep, 105, 329

\bibitem[{{Jang} \& {Lee}(2017{\natexlab{a}})}]{jan17b}
{Jang}, I.~S., \& {Lee}, M.~G. 2017{\natexlab{a}}, \apj, 836, 74

\bibitem[{{Jang} \& {Lee}(2017{\natexlab{b}})}]{jan17a}
---. 2017{\natexlab{b}}, \apj, 835, 28

\bibitem[{{Klotz} \& {Conseil}(2012)}]{klotz_2012}
{Klotz}, A., \& {Conseil}, E. 2012, The Astronomer's Telegram, 4523

\bibitem[{{Komatsu} {et~al.}(2011){Komatsu}, {Smith}, {Dunkley}, {Bennett},
  {Gold}, {Hinshaw}, {Jarosik}, {Larson}, {Nolta}, {Page}, {Spergel},
  {Halpern}, {Hill}, {Kogut}, {Limon}, {Meyer}, {Odegard}, {Tucker}, {Weiland},
  {Wollack}, \& {Wright}}]{kom11}
{Komatsu}, E., {Smith}, K.~M., {Dunkley}, J., {et~al.} 2011, \apjs, 192, 18

\bibitem[{{Krist} {et~al.}(2011){Krist}, {Hook}, \&
  {Stoehr}}]{2011SPIE.8127E..0JK}
{Krist}, J.~E., {Hook}, R.~N., \& {Stoehr}, F. 2011, in \procspie, Vol. 8127,
  Optical Modeling and Performance Predictions V, 81270J

\bibitem[{{Lee} {et~al.}(1993){Lee}, {Freedman}, \& {Madore}}]{lee93}
{Lee}, M.~G., {Freedman}, W.~L., \& {Madore}, B.~F. 1993, \apj, 417, 553

\bibitem[{{Madore} \& {Freedman}(1995)}]{mad95}
{Madore}, B.~F., \& {Freedman}, W.~L. 1995, \aj, 109, 1645

\bibitem[{{Madore} {et~al.}(2009){Madore}, {Mager}, \& {Freedman}}]{mad09}
{Madore}, B.~F., {Mager}, V., \& {Freedman}, W.~L. 2009, \apj, 690, 389

\bibitem[{{Madore} {et~al.}(1998){Madore}, {Freedman}, {Silbermann}, {Harding},
  {Huchra}, {Mould}, {Graham}, {Ferrarese}, {Gibson}, {Han}, {Hoessel},
  {Hughes}, {Illingworth}, {Phelps}, {Sakai}, \& {Stetson}}]{mad98}
{Madore}, B.~F., {Freedman}, W.~L., {Silbermann}, N., {et~al.} 1998, \nat, 395,
  47

\bibitem[{{Madore} {et~al.}(1999){Madore}, {Freedman}, {Silbermann}, {Harding},
  {Huchra}, {Mould}, {Graham}, {Ferrarese}, {Gibson}, {Han}, {Hoessel},
  {Hughes}, {Illingworth}, {Phelps}, {Sakai}, \& {Stetson}}]{mad99}
---. 1999, \apj, 515, 29

\bibitem[{{Mager} {et~al.}(2008){Mager}, {Madore}, \& {Freedman}}]{mag08}
{Mager}, V.~A., {Madore}, B.~F., \& {Freedman}, W.~L. 2008, \apj, 689, 721

\bibitem[{{Makarov} {et~al.}(2006){Makarov}, {Makarova}, {Rizzi}, {Tully},
  {Dolphin}, {Sakai}, \& {Shaya}}]{mak06}
{Makarov}, D., {Makarova}, L., {Rizzi}, L., {et~al.} 2006, \aj, 132, 2729

\bibitem[{{McConnachie} {et~al.}(2004){McConnachie}, {Irwin}, {Ferguson},
  {Ibata}, {Lewis}, \& {Tanvir}}]{mcc04}
{McConnachie}, A.~W., {Irwin}, M.~J., {Ferguson}, A.~M.~N., {et~al.} 2004,
  \mnras, 350, 243

\bibitem[{{M{\'e}ndez} {et~al.}(2002){M{\'e}ndez}, {Davis}, {Moustakas},
  {Newman}, {Madore}, \& {Freedman}}]{men02}
{M{\'e}ndez}, B., {Davis}, M., {Moustakas}, J., {et~al.} 2002, \aj, 124, 213

\bibitem[{{Monson} {et~al.}(2017){Monson}, {Beaton}, {Scowcroft}, {Freedman},
  {Madore}, {Rich}, {Seibert}, {Kollmeier}, \& {Clementini}}]{monson_2017}
{Monson}, A.~J., {Beaton}, R.~L., {Scowcroft}, V., {et~al.} 2017, \aj, 153, 96

\bibitem[{{Mouhcine} {et~al.}(2005){Mouhcine}, {Ferguson}, {Rich}, {Brown}, \&
  {Smith}}]{mou05}
{Mouhcine}, M., {Ferguson}, H.~C., {Rich}, R.~M., {Brown}, T.~M., \& {Smith},
  T.~E. 2005, \apj, 633, 810

\bibitem[{{Persson} {et~al.}(2004){Persson}, {Madore}, {Krzemi{\'n}ski},
  {Freedman}, {Roth}, \& {Murphy}}]{per04}
{Persson}, S.~E., {Madore}, B.~F., {Krzemi{\'n}ski}, W., {et~al.} 2004, \aj,
  128, 2239

\bibitem[{{Persson} {et~al.}(2013){Persson}, {Murphy}, {Smee}, {Birk},
  {Monson}, {Uomoto}, {Koch}, {Shectman}, {Barkhouser}, {Orndorff}, {Hammond},
  {Harding}, {Scharfstein}, {Kelson}, {Marshall}, \& {McCarthy}}]{persson_2013}
{Persson}, S.~E., {Murphy}, D.~C., {Smee}, S., {et~al.} 2013, \pasp, 125, 654

\bibitem[{{Planck Collaboration} {et~al.}(2016){Planck Collaboration}, {Ade},
  {Aghanim}, {Arnaud}, {Ashdown}, {Aumont}, {Baccigalupi}, {Banday},
  {Barreiro}, {Bartlett}, \& et~al.}]{planck16}
{Planck Collaboration}, {Ade}, P.~A.~R., {Aghanim}, N., {et~al.} 2016, \aap,
  594, A13

\bibitem[{{Riess} {et~al.}(2016){Riess}, {Macri}, {Hoffmann}, {Scolnic},
  {Casertano}, {Filippenko}, {Tucker}, {Reid}, {Jones}, {Silverman},
  {Chornock}, {Challis}, {Yuan}, {Brown}, \& {Foley}}]{rie16}
{Riess}, A.~G., {Macri}, L.~M., {Hoffmann}, S.~L., {et~al.} 2016, \apj, 826, 56

\bibitem[{{Rizzi} {et~al.}(2007){Rizzi}, {Tully}, {Makarov}, {Makarova},
  {Dolphin}, {Sakai}, \& {Shaya}}]{riz07}
{Rizzi}, L., {Tully}, R.~B., {Makarov}, D., {et~al.} 2007, \apj, 661, 815

\bibitem[{{Sakai} {et~al.}(1996){Sakai}, {Madore}, \& {Freedman}}]{sak96}
{Sakai}, S., {Madore}, B.~F., \& {Freedman}, W.~L. 1996, \apj, 461, 713

\bibitem[{{Salaris} \& {Cassisi}(1997)}]{salaris_1997}
{Salaris}, M., \& {Cassisi}, S. 1997, \mnras, 289, 406

\bibitem[{{Schlafly} \& {Finkbeiner}(2011)}]{sch11}
{Schlafly}, E.~F., \& {Finkbeiner}, D.~P. 2011, \apj, 737, 103

\bibitem[{{Silbermann} {et~al.}(1999){Silbermann}, {Harding}, {Ferrarese},
  {Stetson}, {Madore}, {Kennicutt}, {Freedman}, {Mould}, {Bresolin}, {Ford},
  {Gibson}, {Graham}, {Han}, {Hoessel}, {Hill}, {Huchra}, {Hughes},
  {Illingworth}, {Kelson}, {Macri}, {Phelps}, {Rawson}, {Sakai}, \&
  {Turner}}]{sil99}
{Silbermann}, N.~A., {Harding}, P., {Ferrarese}, L., {et~al.} 1999, \apj, 515,
  1

\bibitem[{{Sirianni} {et~al.}(2005){Sirianni}, {Jee}, {Ben{\'{\i}}tez},
  {Blakeslee}, {Martel}, {Meurer}, {Clampin}, {De Marchi}, {Ford}, {Gilliland},
  {Hartig}, {Illingworth}, {Mack}, \& {McCann}}]{2005PASP..117.1049S}
{Sirianni}, M., {Jee}, M.~J., {Ben{\'{\i}}tez}, N., {et~al.} 2005, \pasp, 117,
  1049

\bibitem[{{Stetson}(1987)}]{1987PASP...99..191S}
{Stetson}, P.~B. 1987, \pasp, 99, 191

\bibitem[{{Stetson}(1994)}]{1994PASP..106..250S}
---. 1994, \pasp, 106, 250

\bibitem[{{Turner}(1995)}]{allframecookbook}
{Turner}, A.~M. 1995

\end{thebibliography}

\appendix 

\section{Comparison of \cchp and Literature Techniques}\label{app} 

We undertake comparisons at three stages of the data reduction and analysis: 
(i) image processing and photometry (Section \ref{app:redux}),
(ii) rectification of the color-sensitivity of the RGB (Section \ref{app:redux}), and 
(iii) testing of other edge-detection techniques (Section \ref{app:edge}).

\subsection{Comparison of FLC and drizzled photometry} \label{app:redux} 
The RGB stars measured in this study are as faint as F814W $\approx$ 28 mag and F606W $\approx$ 29 mag.
In individual frames (20 for F814W and 12 or F606W), these stars are measured at low signal to noise.
There are two independent approaches for producing photometry for these sources:
\begin{enumerate}
   \item Generate a master source list from a high $S/N$ median image and use it as an input to force-photometer individual frames (as is done in the {\sc ALLFRAME} software). The photometry is completed on the flc image products and we will refer to this technique as {\sc flc}.
   \item Directly photometer co-added images, defining an empirical PSF based on high $S/N$ sources in the median image. The photometry is completed on a drc image product and we will refer to this technique as {\sc drc}.
\end{enumerate}
The former ({\sc flc}) is the technique described in the main text, for which we utilize the theoretical Tiny Tim PSFs \citep{2011SPIE.8127E..0JK}. It has the disadvantage of the stellar full-width-at-half-maximum ({\sc fwhm}) being under-sampled (though we note that because stellar crowding is low, our stellar profile fitting is not limited to the stellar {\sc fwhm}).
The latter technique ({\sc drc}) has been used more broadly in the literature for TRGB-based analyses at these distances \citep[e.g.,][and references therein]{jan17b} and comes with the advantage of producing stellar profiles that are Nyquist sampled within the stellar {\sc fwhm}. 
In this Section, we provide quantitative comparisons between the {\sc flc} and {\sc drc} methods.

We adopt the {\sc flc} photometry from the main text and the {\sc drc} photometry is produced as follows. 
Drizzled image stacks are constructed using {\tt DrizzlePac} \citep{fru02}. 
We carefully selected $\sim$100 relatively bright sources in each CCD chip and used them to refine image alignment with the {\tt Tweakreg} task; the mean residual RMS for the $X$ and $Y$ shifts determined with {\tt Tweakreg} were smaller than 0.1 pixel. We then used {\tt Astrodrizzle} to make a combined drizzled image for each filter with {\tt final\_pixfrac} = 0.8 and {\tt final\_scale} = 0.03 arcsec pixel$^{-1}$. The output drizzled images have stellar {\sc FWHM}s of $\sim$3 pixels, corresponding to $\sim$0.09$\arcsec$. PSF photometry on the drizzled images and standard calibration were performed following the method described in main text with the exception of the PSF modeling. We generated empirical PSFs with {\sc DAOPHOT} that were constructed from $\sim15$ bright isolated stars in each of the F606W and F814W images. The F814W source catalog is used as the `master catalog' and the two frames are simultaneously photometered in {\sc ALLFRAME} \citep{1994PASP..106..250S}. The magnitudes are calibrated identically as described in the main text.

\begin{figure*} 
\centering
\includegraphics[width=0.4\textwidth]{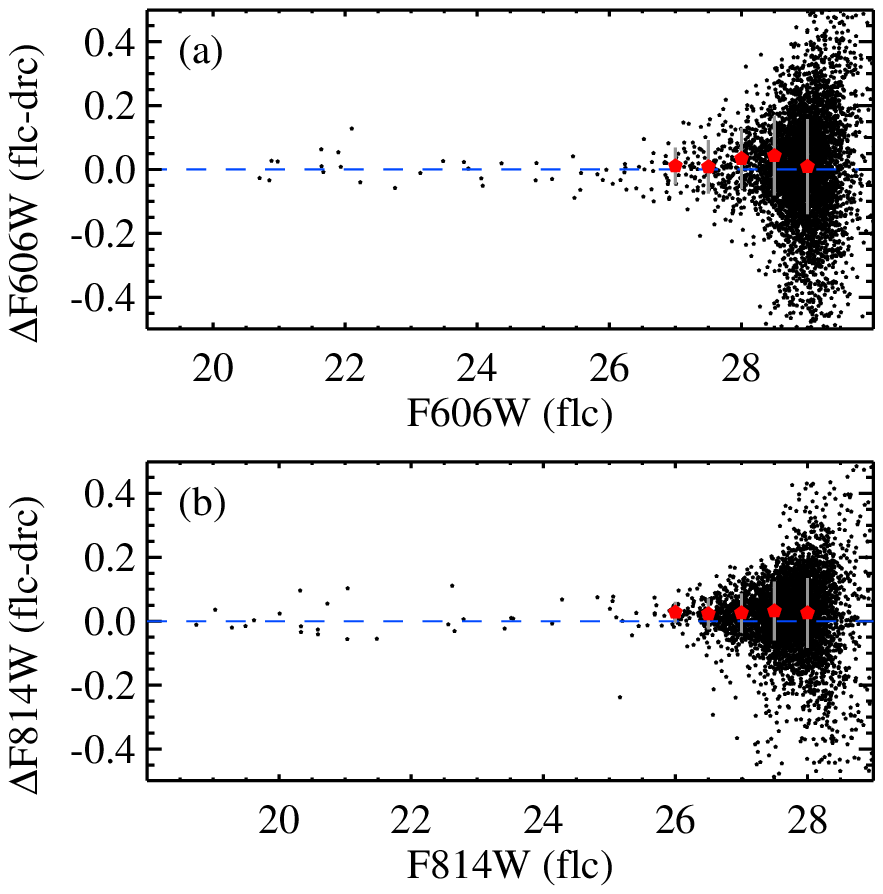}
\includegraphics[width=0.4\textwidth]{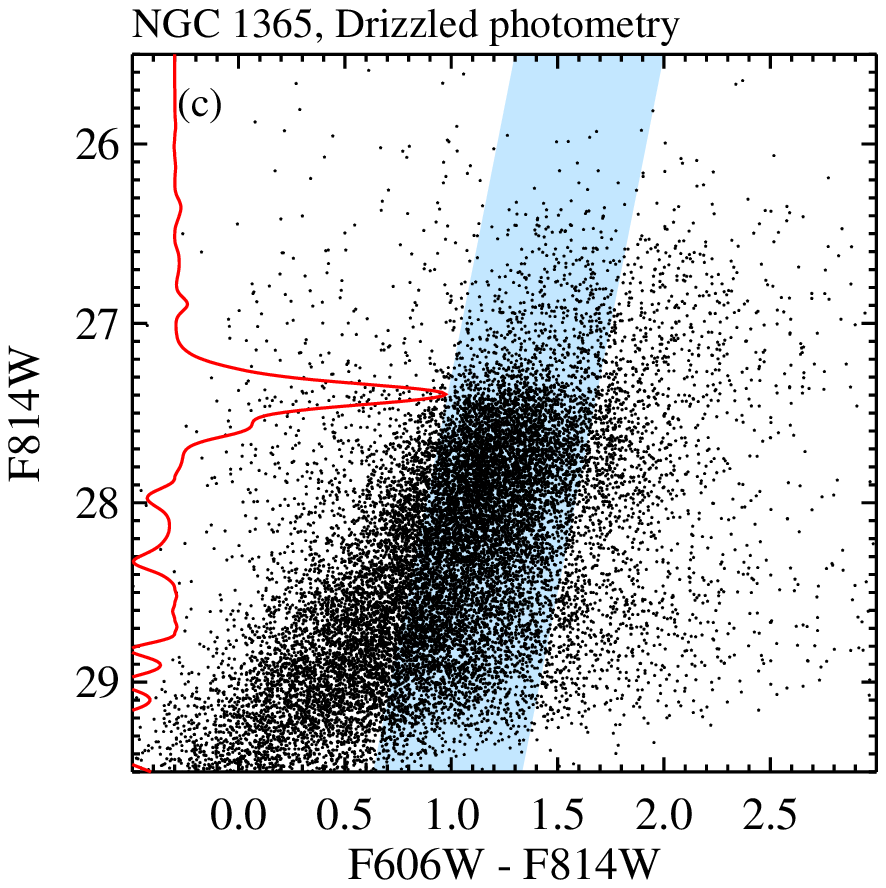}
\caption{ \label{fig:drizzle}
Comparison of PSF magnitudes in F606W (a) and F814W (b) between individual frame photometry with TinyTim PSFs (flc) and drizzled photometry with empirical PSFs (drc). Median offsets and standard deviations at each magnitude bin are marked by red pentagons and error bars, respectively. 
(c) F814W -- (F606W-F814W) CMD of resolved stars in NGC 1365 from drizzled photometry (drc). 
An edge detection response (solid line) for stars in the shaded region shows a strong peak at F814W $=27.39\pm0.03$, which is statistically identical to that determined in the main text.} 
\end{figure*}

Figures \ref{fig:drizzle}a and \ref{fig:drizzle}b provide star-by-star comparisons of the {\sc flc} and {\sc drc} photometry in the F606W and F814W filters, respectively. Bright stars with F606W $\lesssim 24$ mag and F814W $\lesssim 23$ mag are in excellent agreement with median offsets smaller than 0.01 mag for both filters. However, we measure small systematic offsets for the fainter stars. At the TRGB magnitude (F814W$\approx$ 27.4 mag and F606W $\approx$ 28.7 mag), median offsets are measured to be 0.04 mag in F606W and 0.03 mag in F814W. The precise origin of the offsets for fainter sources remains unclear, but could be due to (i) the relatively small number of sources used to determine the empirical PSF\footnote{Using a small number of sources limits the ability of the {\sc DAOPHOT}-based PSF model to properly account for the PSF variation across the frame due to residual distortion or true variation. Moreover, the PSF is more susceptible to non-stellar contaminants and other non-ideal features in the profile.} or (ii) documented differences between the Tiny Tim and empirical magnitudes for faint sources that were described by \citet{2011SPIE.8127E..0JK}. For 282 sources between 27.34 \textless ~F814W \textless ~27.40 mag the median magnitude uncertainty is 0.068 mag for F814W and 0.13 mag for F606W (the latter measurement is for the same stars in the F814W range). Thus, the differences identified in Figures \ref{fig:drizzle}a and \ref{fig:drizzle}b for the fainter sources are within the magnitude uncertainties.

While some star-to-star differences are demonstrated in Figures \ref{fig:drizzle}a and \ref{fig:drizzle}b, a more relevant question is the results from the TRGB detection. Thus, we employ the same techniques described in the main text to the {\sc drc} photometry. The result is given in Figure \ref{fig:drizzle}c. The CMD shows a well defined RGB with a visible discontinuity TRGB at F814W $\sim$ 27.4 mag. A visual comparison to Figure \ref{fig:f2} reveals that the drizzle based CMD looks more well populated (i.e., more complete) and this is consistent with having performed PSF photometry on a higher signal-to-noise image. We selected stars in the shaded region (identical to that of Figure \ref{fig:f2} in the main text), construct a luminosity function, apply the GLOESS smoothing, and, lastly, apply the [-1,0,1] Sobel filter. The edge detection response is shown in red in Figure \ref{fig:drizzle}c and the maximal response is $=27.39\pm0.03$ mag. Derived TRGB magnitude is statistically consistent with the value from the individual frame photometry, F814W = \trgbobsval~ $\pm$ \trgbobsvalstaterr ~mag (Table \ref{tab_distance}). 

From the comparisons given in the panels of Figure \ref{fig:drizzle}, we conclude that the two reduction procedures are statistically identical for both their output photometry and in their TRGB measurements. Thus, the choice to use individual frame photometry for the \cchp project, motivated by the need for a homogeneous image processing strategy for both nearby and distant \sne Ia hosts, is consistent with the body of work derived from drizzled photometry \citep[e.g.,][]{jan17b}.

\begin{figure} 
\centering
\includegraphics[width=0.7\textwidth]{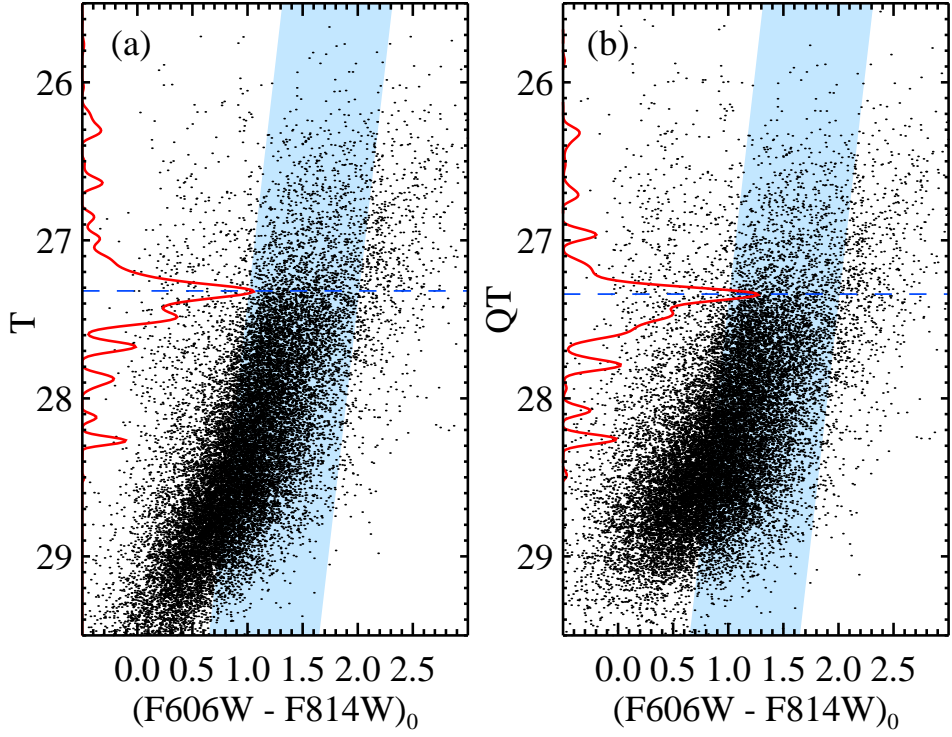} 
\caption{\ngc color-magnitude diagrams using the rectified TRGB method in the (a) $T_{F814W,F606W}$ and (b) $QT_{F814W,F606W}$ systems described in the text. The edge detection response from the Sobel filter, $[-1, 0, +1]$ applied to the GLOESS smoothed LF is shown by red solid lines in each panel.
}
\label{fig:fb1}
\end{figure} 

\subsection{Rectified TRGB Magnitudes}\label{app:retified} 

A great benefit of the TRGB as a distance indicator is that the metallicity sensitivity of the absolute magnitude is projected into the color of the star. Furthermore, for metal-poor stars that populate the `blue' edge of the TRGB, $(V-I)_0\lesssim2$, the $I$ magnitude of the TRGB is relatively insensitive to metallicity \citep[i.e., is flat with color; see][among others]{lee93, riz07, mad09, jan17a}. Thus, with only an optical color-cut (as is done in Figure \ref{fig:f2}), the $I$, and by proxy the F814W TRGB, requires no correction for metallicity to convert to an absolute magnitude system.

The color dependence of the $I$ TRGB, however, is not everywhere negligible; in particular, for the color range $(V-I)_0 \gtrsim 2.0$ the $I$ magnitude becomes noticeably fainter. 
\citet{mad09} presented an empirical technique to rectify or transform the TRGB magnitudes for metal-rich sources to the metal-poor (flat) portion of the TRGB. 
The general form of a transformation into the $T_{\lambda_{1},\lambda_{2}}$ (or TRGB) magnitude system is defined as
\begin{equation}
T_{\lambda_{1},\lambda_{2}}=m_{\lambda_{1}} - \beta_{\lambda_{1},\lambda_{2}} [(m_{\lambda_{1}}-m_{\lambda_{2}})-\gamma_{\lambda_{1},\lambda_{2}}]  
\end{equation}
where the $T_{\lambda_{1},\lambda_{2}}$ is the initial magnitude ($m_{\lambda_{1}}$) corrected for the slope of the TRGB ($\beta_{\lambda_{1},\lambda_{2}}$) to a fiducial color $\gamma_{\lambda_{1},\lambda_{2}}$.
In the standard Johnson-Cousins system used in \citet{mad09}, the slope is $\beta_{I,V} = 0.2$ and the fiducial color was $\gamma_{I,V} = 1.5$ to produce $T_{I,V}$ magnitudes from $I$ photometry. The parameter values were  determined from a linear approximation to the TRGB predicted by theoretical models described by \citet{bel01, bel04}. A standard edge-detection algorithm can be applied to the rectified CMD and the distance modulus is computed as $(m-M)_0=T-M_{TRGB}$, where $M_{TRGB}$ is defined at the fiducial color, $\gamma_{\lambda_{1},\lambda_{2}}$.

For use in this work, we convert the \citet{mad09} $T$ magnitude system into the ACS/WFC system for $\lambda_{1}$=F814W with $\lambda_{2}$=F606W and $\lambda_{2}$=F555W utilizing the photometric transformations from the flight magnitude system to the Johnson-Cousins system given in \cite{2005PASP..117.1049S}. 
We find $\beta_{F814W,F606W} = 0.27$ mag color$^{-1}$ and the fiducial color is $\gamma_{F814W,F606W} = 1.18$ mag and $\beta_{F814W,F555W} = 0.19$ mag color$^{-1}$ and the fiducial color is $\gamma_{F814W,F555W} = 1.59$ mag.
These conversions are approximate only and should be measured directly from color-magnitude diagrams in these filters.

\citet{jan17a} investigated the color dependence of the TRGB from the HST/ACS photometry of eight nearby galaxies and find that the run of the $I$ TRGB with the $V-I$ color can be described with two components: a flat one for the blue color range ($V-I\lesssim1.9$) and a steep component for the red color range ($V-I\gtrsim1.9$). From this, they introduced the $QT$ magnitude, a quadratic form of the TRGB magnitude corrected for the color dependence of the TRGB. $QT_{\lambda_{1},\lambda_{2}}$ is given by
 \begin{equation}
    QT_{\lambda_{1},\lambda_{2}}=m_{\lambda_{1}} - \beta_{\lambda_{1},\lambda_{2}} [(m_{\lambda_{1}}-m_{\lambda_{2}})-\gamma_{\lambda_{1},\lambda_{2}}] - \alpha_{\lambda_{1},\lambda_{2}} [(m_{\lambda_{1}}-m_{\lambda_{2}})-\gamma_{\lambda_{1},\lambda_{2}}]^2 \\
 \end{equation}
where $\alpha_{F814W,F606W}=0.159\pm0.010$ mag color$^{-2}$, $\beta_{F814W,F606W}=-0.047\pm0.020$ mag color$^{-1}$, and $\gamma_{F814W,F606W}=1.1$ mag. 

We applied the $T_{F814W,F606W}$ and $QT_{F814W,F606W}$ magnitude transformations to the \ngc~{\sc flc} photometry and the resulting color-magnitude diagrams are shown in Figures \ref{fig:fb1}a and \ref{fig:fb1}b, respectively.
To construct a LF, we apply the color-magnitude restriction indicated by the blue shading in Figures \ref{fig:fb1}a and \ref{fig:fb1}b, which is identical to that applied in Figures \ref{fig:f2} and \ref{fig:n1365trgbmeas}a in the main text.
We use GLOESS smoothing with our idealized \sigmasmooth and use the  Sobel filter, $[-1, 0, +1]$. 
The edge-detection response function is shown in Figures \ref{fig:fb1}a and \ref{fig:fb1}b and has strong peaks at $T\simeq QT\simeq 27.4$ mag.
The TRGB magnitude and uncertainties are derived following the procedure outlined in \citet{jan17b}, which uses the results of bootstrap re-sampling to define the true TRGB tip and its uncertainty.
We obtain TRGB magnitudes: $T_{F814W,F606W} = 27.32\pm0.03$ mag and $QT_{F814W,F606W} = 27.34\pm0.03$ mag, which agree within their mutual uncertainties. 

Comparing the $T_{F814W,F606W}$ and $QT_{F814W,F606W}$ results to that from the main text, F814W = \trgbobsval~ $\pm$ \trgbobsvalstaterr ~mag (Table \ref{tab_distance}), we find agreement within the quoted uncertainties. We note that the absolute magnitude of the TRGB is shifted systematically fainter at the $\sim$0.01 mag level for the $T_{F814W,F606W}$ and $QT_{F814W,F606W}$ systems, which also brings the measurements into better agreement \citep[we refer the reader to][for details]{jan17a}.
As was mentioned in the previous section, the median magnitude uncertainty is 0.068 mag for F814W and 0.13 mag for F606W at the TRGB, which means that the  $T_{F814W,F606W}$ and $QT_{F814W,F606W}$ magnitudes, themselves, are at significantly larger uncertainties and will scatter (preferentially brighter in the form of the transformation). This is visually apparent by comparing Figure \ref{fig:f2} to Figures \ref{fig:fb1}a and \ref{fig:fb1}b; in particular, the visible density of stars near the tip does not look to be significantly improved by moving into the $T_{F814W,F606W}$ and $QT_{F814W,F606W}$ systems. We further note that our color-magnitude restrictions largely avoid the regions of F606W-F814W color where the $T_{F814W,F606W}$ and $QT_{F814W,F606W}$ magnitudes are expected to provide the most benefit by bringing these fainter TRGB sources to the same magnitude of the bluer TRGB. Thus, we conclude that our results from the raw F814W magnitudes are fully consistent with those determined with the rectified $T_{F814W,F606W}$ and $QT_{F814W,F606W}$ systems. 

\begin{figure*} 
\centering
\includegraphics[width=0.75\textwidth]{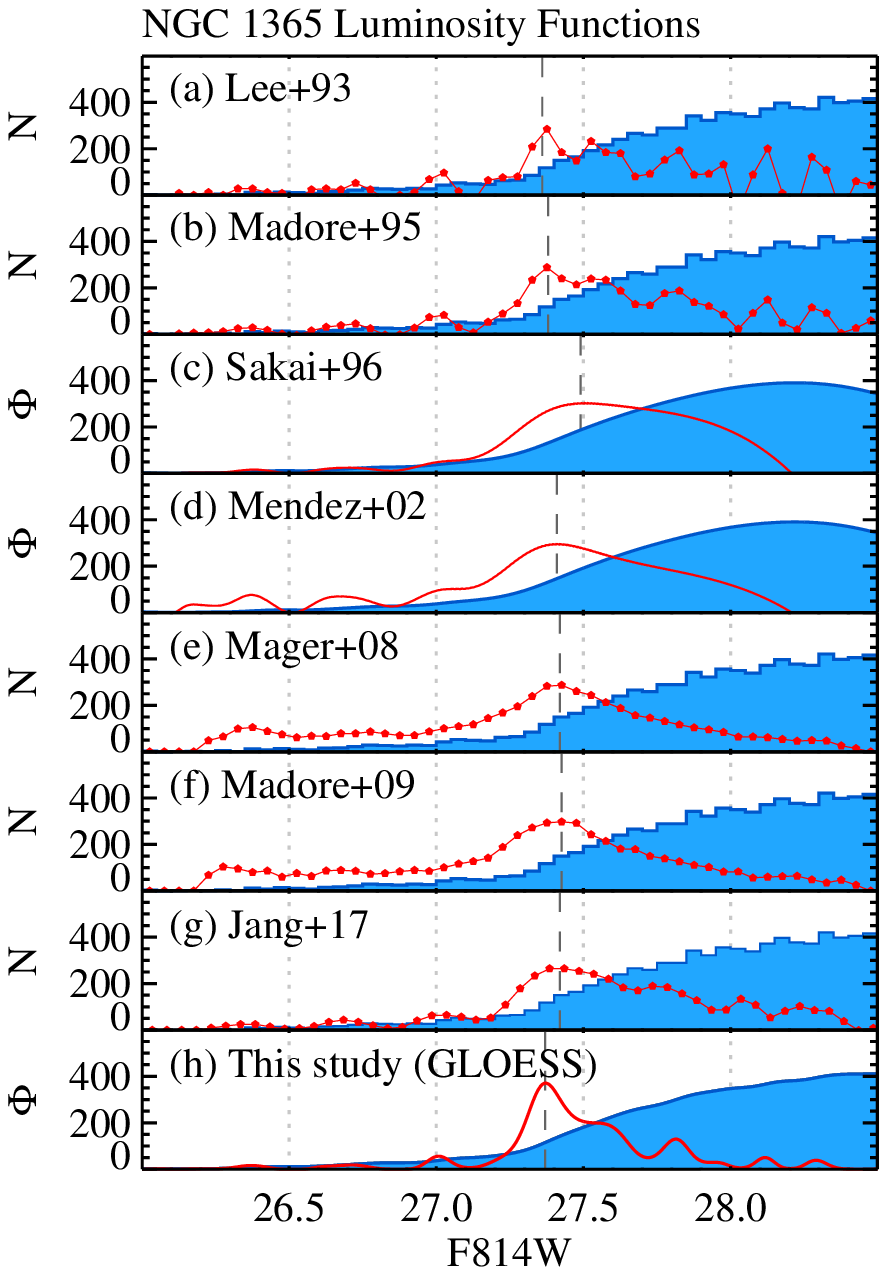}
 \caption{F814W luminosity functions for RGB stars in \ngc (blue histograms) and edge detection responses (red lines) obtained from several edge detection algorithms: (a) \citet{lee93}, (b) \citet{mad95}, (c) \citet{sak96}, (d) \citet{men02}, (e) \citet{mag08}, (f) \citet{mad09}, (g) \citet{jan17b}, and (h) this study. The histograms are modified in each panel as per the prescription for the edge-detection algorithm employed with $N$ representing binned LFs and $\phi$ representing LFs modified with smoothing. The TRGB magnitude as determined by the edge-detector run on the LF is marked by a dashed line in each panel.
 } 
 \label{fig:fc1} 
 \end{figure*} 

\subsection{Comparison of Edge Detectors}\label{app:edge} 

The detection of the apparent magnitude of the TRGB is one of the most critical steps in the TRGB distance estimation. 
Broadly, two independent approaches have been developed for identifying the TRGB: 
\begin{enumerate} 
\item Direct edge detection algorithms as in \citet{lee93, mad95, sak96, men02, mag08, mad09} that typically make use of a form of the Sobel Filter that is an approximation to the first derivative of a discrete function. These can take on discrete ($N$) and continuous forms ($\phi$) based on the smoothing that is applied to the LF before application of the edge detection algorithm. 
\item Template fitting as in \citet{cio00, men02, fra03, mcc04, mou05, mak06, con11}.
\end{enumerate}
Based on a review of the literature, we have selected seven forms of TRGB edge detection in addition to that adopted by the \cchp. These are similar to those applied in Appendix B of \citetalias{hatt17}.

We have applied the eight Sobel filters to the luminosity functions of the selected stars in NGC 1365 and plot them in Figure \ref{fig:fc1}. 
We used a 0.05 mag bin to construct the LF for those edge-detectors that operate directly on the histogram (e.g., Figures \ref{fig:fc1}a, \ref{fig:fc1}b, \ref{fig:fc1}e, \ref{fig:fc1}f, and \ref{fig:fc1}g). 
In the case of the continuous forms of Sobel filters, we used a bin width of 0.001 mag for deriving the Gaussian smoothed luminosity functions (e.g., Figures \ref{fig:fc1}c and \ref{fig:fc1}d). 
Figure \ref{fig:fc1}h is a re-visualization of the algorithm applied in the main text. 
The magnitude of the TRGB are determined by choosing the maximum edge-detection response in Figures \ref{fig:fc1}c, \ref{fig:fc1}d, and \ref{fig:fc1}h) and via the \citet{jan17b} bootstrap resampling method in Figures \ref{fig:fc1}a, \ref{fig:fc1}b, \ref{fig:fc1}e, \ref{fig:fc1}f, and \ref{fig:fc1}g.
Qualitatively, all eight edge-detection responses in the panels of Figure \ref{fig:fc1} have peaks at F814W$\sim27.4$ mag.

The results of the quantitative tip detection are as follows in the panels of Figure \ref{fig:fc1} as follows:
in Figure \ref{fig:fc1}a the TRGB = 27.36 mag from the \citet{lee93} algorithm,
in Figure \ref{fig:fc1}b the TRGB = 27.38 mag from the \citet{mad95} algorithm,
in Figure \ref{fig:fc1}c the TRGB = 27.49 mag from the \citet{sak96} algorithm,
in Figure \ref{fig:fc1}d the TRGB = 27.41 mag from the \citet{men02} algorithm,
in Figure \ref{fig:fc1}e the TRGB = 27.42 mag from the \citet{mag08} algorithm,
in Figure \ref{fig:fc1}f the TRGB = 27.43 mag from the \citet{mad09} algorithm,
in Figure \ref{fig:fc1}g the TRGB = 27.42 mag from the \citet{jan17b} algorithm,
in Figure \ref{fig:fc1}h the TRGB = 27.37 mag from the algorithm adopted in the main text.
In each panel of Figure \ref{fig:fc1}, the TRGB is indicated by the vertical dashed line in each panel.
While there is qualitative agreement, the the techniques produce results that vary over a range of 0.13 mag, which is four times larger than the quoted uncertainty on our measurement of 0.03 mag. 
In \citetalias{hatt17}, a thorough discussion of the advantages and disadvantages of the wide-range of edge detectors was presented and for brevity we will only discuss the implications that can be interpreted from the panels of Figure \ref{fig:fc1}.

First, the use of large bins is problematic since not only the size of the bin, but also the starting point of the bins have an effect on the quantitative Sobel response. Thus, in our discretely binned LFs, an additional random and systematic uncertainty of $\sim$0.03 (50\% of a bin) should be added to the algorithmic measurement uncertainty (e.g., the $\sim$0.03 mag uncertainty derived from bootstrap resampling). These two components arise due to the inability to distinguish the location of the peak within the set binning strategy and this must be applied to the results in Figures \ref{fig:fc1}a, \ref{fig:fc1}b, \ref{fig:fc1}e, \ref{fig:fc1}f, and \ref{fig:fc1}g. Allowing for these additional uncertainties, all of these values are consistent with our measurement (Figure \ref{fig:fc1}h).

Second, many of the edge-detection algorithms themselves employ smoothing directly into the algorithm itself. If applied to a `raw' LF, then this is not problematic, but many of the algorithms are applied to LFs that have already been smoothed. This is clearly evident in Figure \ref{fig:fc1}c, which has not only a heavily smoothed LF, but also a heavily smoothed algorithm. This `double smoothing' results in the most deviant of the TRGB values (27.49 mag) and the response of the edge-detection is not a peak, but a plateau that reduces the precision of the tool. The bias in the \citet{sak96} algorithm is evident by comparing Figures \ref{fig:fc1}c and \ref{fig:fc1}d, that while having nearly identical LFs, have very different edge responses. 

Lastly, there are algorithms that attempt to model the uncertainties in the data (both magnitude uncertainties and completeness), but these rely critically on the ability to assess these values well for a dataset. After being modeled, these uncertainties are folded into the detection algorithm itself, instead of applying modifications to the LF directly. The difficulty with this approach is that it is not fully reproducible by an independent team. As has been shown in previous sections of this Appendix, there are quantitative differences at the 0.04 mag level between photometry derived from the same underlying images due to subtle choices in the data processing. We have demonstrated that our algorithms for the LF and for the edge-detection are robust to these differences, but algorithmic approaches that use the photometry characterizations directly from one's own photometry would not be reproducible by an independent process. This is particularly concerning for the template-fitting strategies that rely on (i) the input idealized model of the LF to be well matched to the  actual intrinsic luminosity function for the field of interest and (ii) are highly sensitive to the completeness in both bands of the photometry (not just the band used for the LF). 

In conclusion, we see quantitative differences between our adopted strategy for smoothing the LF and for applying an edge-detection algorithm (Figure \ref{fig:fc1}). As we have shown, these differences can be understood within the true uncertainties of the various techniques. As discussed in depth in \citetalias{hatt17}, our LF binning, edge-detection algorithm, and our modeling of the uncertainties are explicitly designed to be reproducible by others and to take into account the full scale photometric uncertainties. 

\end{document}